\documentclass[twocolumn,eqsecnum,showpacs,preprintnumbers,amsmath,amssymb,nofootinbib,english]{revtex4}

\usepackage{babel}
\usepackage{dcolumn}% Align table columns on decimal point

\newcommand{\grad}{\mathrm{d}}
\newcommand{\rot}{\mathrm{rot}\,}
\renewcommand{\div}{\mathrm{div}\,}
\newcommand{\laplace}{{\vartriangle}\,}

\newcommand{\eps}{\varepsilon}
\newcommand{\epso}{\varepsilon_{\mathrm{o}}}
\newcommand{\vrho}{\varrho_{\EM}}
\newcommand{\sslfdl}{s}
\newcommand{\sreal}{\sigma}

\newcommand{\frm}[1]{{\Theta^{(#1)}}}
\newcommand{\mb}{{\overline{m}}}
\newcommand{\trans}{{\scriptscriptstyle\top}}

\newcommand{\NP}[1]{{#1}_{\scriptscriptstyle\mathrm{NP}}}
\newcommand{\EM}{{\scriptscriptstyle\mathrm{EM}}}
\newcommand{\gyr}{{\scriptscriptstyle\mathrm{gyr}}}
\newcommand{\ro}{{\mathrm{o}}}

\begin{document}

\title{Gyratons on Melvin spacetime}

\author{Hedvika Kadlecov\'{a}}
\email{hedvika.kadlecova@centrum.cz}%
\affiliation{
Institute of Theoretical Physics, Faculty of Mathematics and Physics, Charles University,\\
  V Hole\v{s}ovi\v{c}k\'{a}ch 2, 180 00 Prague 8, Czech Republic
 }

\author{Pavel Krtou\v{s}}
\email{Pavel.Krtous@mff.cuni.cz}%
\affiliation{
Institute of Theoretical Physics, Faculty of Mathematics and Physics, Charles University,\\
  V Hole\v{s}ovi\v{c}k\'{a}ch 2, 180 00 Prague 8, Czech Republic
  }

\date{June 9, 2010}% version 2.02 [arxiv v3]

\begin{abstract}{We present and analyze new exact gyraton solutions of algebraic type II on a background
 which is a static, cylindrically symmetric Melvin universe of type D. For a vanishing
 electromagnetic field it reduces to previously studied gyratons on Minkowski background.
 We demonstrate that the solutions are a member of a more general family of the Kundt spacetimes.
 We show that the Einstein equations reduce to a set of mostly linear equations on a transverse 2-space
 and we discuss the properties of polynomial scalar curvature invariants which are generally non--constant
 but unaffected by the presence of gyratons.}
\end{abstract}

% 04.20.Jb Exact solutions
% 04.30.-w Gravitational waves: theory
% 04.40.Nr Einstein--Maxwell spacetimes, spacetimes with fluids, radiation or classical fields

\maketitle

%%%%%%%%%%%%%%%%%%%%%%%%%%%%%%%%%%%%%%%%%%%%%%%%%%%%%%%%%%%%%%%%%%%%%%%%%%%%%%%%%%%%%
\section{\label{sec:level1}Introduction}
%%%%%%%%%%%%%%%%%%%%%%%%%%%%%%%%%%%%%%%%%%%%%%%%%%%%%%%%%%%%%%%%%%%%%%%%%%%%%%%%%%%%%%%

Gyraton solutions represent the gravitational field of
a localized matter source with an intrinsic rotation
which is moving at the speed of light.
Such an idealized ultrarelativistic source can describe
a pulse of a spinning radiation beam and it is accompanied
by a sandwich or impulsive gravitational wave.

The gravitational fields generated by (nonrotating) light
pulses and beams were already studied by Tolman \cite{Tol:1934:Oxf:}
in 1934, who obtained the corresponding solution in the
linear approximation of the Einstein theory.
Exact solutions of the Einstein--Maxwell equations for such `pencils of light'
were found and analyzed by Peres \cite{Peres:1960:PHYSR:} and Bonnor
\cite{Bonnor:1969:COMMPH:,Bonnor:1969:INTHP:,Bonnor:1970a:INTHP:}.
These solutions belong to a general family of {\it pp\,}-waves
\cite{Step:2003:Cam:,GrifPod:2009:Cam:}.

In the impulsive limit (i.e., for an infinitely thin beam,
and for the delta-type distribution of the light-pulse in time),
the simplest of these solutions represents the well-known
Aichelburg--Sexl metric \cite{Aich-Sexl:1971:} which describes
the field of a pointlike null particle.  Subsequently,
more general impulsive waves were found
\cite{FerPen90,LouSan92,HotTan93,KBalNac95,KBalNac96,PodGri97,PodGri98prd}
(for recent reviews see \cite{Podolsky02,BarHog:2003:WorldSci:}).

The gyraton solutions are generalization of
{\it pp\,}-waves which belong to the Kundt class
for which the source---the beam of radiation---carries not only energy,
but also an additional angular momentum. Such spacetimes were
first considered by Bonnor in \cite{Bonnor:1970b:INTHP:},
who studied the gravitational field created by a spinning null fluid.
In some cases, this may be interpreted as a
massless neutrino field~\cite{Griffiths:1972a:INTHP:}.

Gyratons on Minkowski background are locally isometric to standard
{\it pp\,}-waves in the exterior vacuum region, outside the source.
The interior region contains a nonexpanding null matter which possesses
an intrinsic spin. In general, these solutions are obtained by keeping
nondiagonal terms $g_{ui}$ in the Brinkmann form \cite{Brink:1925:MAAN:}
of the {\it pp\,}-wave solution, where $u$ is the null coordinate
and $x^i$ are orthogonal spatial coordinates. The corresponding
energy-momentum tensor thus also contains an extra nondiagonal term
$T_{ui}=j_{i}$. In four dimensions, the terms $g_{ui}$ can be set
to zero {\it locally}, using a suitable gauge transformation.
However, they can not be {\it globally} removed because the gauge
invariant contour integral $\oint g_{ui}(u,x^{j})\,\grad x^{i}$
around the position of the gyraton is proportional to the nonzero
angular momentum density $j_{i}$, which is nonvanishing.

These gyratons were investigated (in the linear approximation)
in \cite{Fro-Fur:2005:PHYSR4:} in higher dimensional flat space,
the exact gyraton solutions propagating in an asymptotically
flat $D$-dimensional spacetime were further investigated
in \cite{Fro-Is-Zel:2005:PHYSR4:}. They proved that the Einstein's
equations for gyratons reduce to a set of linear equations
in the Euclidean ${(D-2)}$-dimensional space and showed that
the gyraton metrics belong to a class of so called \emph{VSI spacetimes}
for which all polynomial scalar invariants, constructed from the curvature
and its covariant derivatives, vanish identically
\cite{Prav-Prav:2002:CLAQG:}. (For the discussion of spacetimes
with nonvanishing but nonpolynomial scalar invariants of curvature,
see \cite{Page:2009:}.) Subsequently, charged gyratons in Minkowski
space in any dimension were presented in \cite{Fro-Zel:2006:CLAQG:}.

In \cite{Fro-Zel:2005:PHYSR4:}, the exact gyraton solutions
in the asymptotically anti-de~Sitter spacetime were found.
Namely, they obtained Siklos gyratons which generalize the
Siklos family of nonexpanding waves \cite{Sik:1985:Cam:}
(investigated further in \cite{Pod-rot:1998:CLAQG:})
which belong to the class of spacetimes with constant scalar invariants
(\emph{CSI spacetimes}) \cite{Coley-Her-Pel:2006:CLAQG, Coley-Gib-Her-Pope:2008:CLAQG, Coley-Her-Pel:2008:CLAQG, Coley-Her-Pel:2009:CLAQG}.

Recently, the large class of gyratons on the direct-product
spacetimes was found in \cite{Kadlecova:2009:PHYSR4:}, where we showed that
this class of gyratons has similar properties as the previous
gyratonic solutions: the Einstein's equations reduce to a set of
linear equations in transversal 2-space and these spacetimes
belong to the CSI class of spacetimes.

Let us also mention that string gyratons in supergravity
were recently found in \cite{Fro-Li:2006:PHYSR4:}.
Supersymmetric gyraton solutions were also obtained in minimal gauged theory
in five dimensions in
\cite{Cald-Kle-Zor:2007:CLAQG:}, where the configuration represents
a generalization of the Siklos waves with a nonzero
angular momentum in anti-de~Sitter space.

The gyratons are important in studies of production of mini
black holes or in cosmic ray experiments.
The theory of high energy particle collisions was developed in
\cite{Yosh:2005:PHYSR4:,Yosh:2006:PHYSR4:}
and was applied to gyraton models in \cite{Yosh:2007:PHYSR4:}.

The main purpose of this paper is to further extend the family of
gyratonic solutions. In particular, we present new gyraton solutions
of algebraic type~II, propagating in the Melvin universe.

The Melvin universe \cite{Bonnor:1954:PRS:,Melvin:1965:PHYSR:} is a nonsingular
electro-vacuum solution with physical properties which are interesting
both from a classical and a quantum point of view. The spacetime represents
a parallel bundle of magnetic (or electric) flux held together by its
own gravitational attraction. The transverse space orthogonal to the direction of the flux
has a nontrivial spatial geometry. It was represented in \cite{Thorne:1965:PHYSR:}
by a suitable embedding diagram which resembles a tall narrow-necked vase.
Also in \cite{Thorne:1965:PHYSR:,MelvinWallingford:1966:JMATHP:} it was shown
that no motion can get too far from the axis of symmetry.
This aspect is analogous to the attractive effect of a negative cosmological
constant in the anti-de Sitter universe.

The Melvin universe was considered as
an important model in astrophysical processes related to gravitational
collapse because of its stability. It was shown in \cite{Melvin:1965:PHYSR:,Thorne:1965:PHYSR:}
that the spacetime is surprisingly stable against small radial perturbations
and also against large perturbations which are concentrated in
a finite region about the axis of symmetry. The asymmetries are radiated
away in gravitational and electromagnetic waves
\cite{GarfinkleMelvin:1992:PHYSR4:,Ortaggio:2004:PHYSR4:}.

The Melvin universe also appears as a limit in more
complicated solutions, in \cite{HavrdovaKrtous:2007:GENRG2:} it is obtained
as a specific limit of a charged C-metric.
The Melvin universe has been generalized to Kaluza-Klein and dilaton
theories \cite{GibbonsMaeda:1988:NUCLB:}, to nonlinear electrodynamics
\cite{GibbonsHerdeiro:2001:CLAQG:}, and has important applications
in the study of quantum black hole pair creation in a background electromagnetic field
\cite{Gib:1986:Sin:,GarfinkleStrominger:1991:PHRELEB:,GarfinkleGidStro:1994:PHYSR4:,%
DowkerGauntKaTra:1994:PHYSR4:,DowkerGauntGiddHorowitz:1994:PHYSR4:,%
HawkingRoss:1995:PHYSR4:,Emparan:1995:PHRELE:}.

The Melvin universe recently attracted a new interest because it is possible
to find gravitational waves in the Melvin universe
\cite{GarfinkleMelvin:1992:PHYSR4:} by an ultrarelativistic boost of the
Schwarzschild--Melvin black hole metric \cite{Ortaggio:2004:PHYSR4:}.
It was shown that these wave solutions are straightforward impulsive
limits of a more general class of Kundt spacetimes of type II with
an arbitrary profile function, which can be interpreted as gravitational
waves propagating on the Melvin spacetime. The gyraton spacetimes
investigated in this paper are generalizations of such Kundt waves
when their ultrarelativistic source is made of a `spinning matter'.

%%%%%%%%%%%%%%%%%%%%%%%%%%%%%%%%%%%%%%%%%%%%%%%%%%%%%%%%%%%%%%%%%%%%%%%%%%%%%%%%%%%%%%%%%

The paper is organized as follows. In Section \ref{sc:gyreq} we review
basic information about the Melvin universe which will be useful in the paper.
We derive the ansatz for the gyraton metric by a direct transformation
from the Kundt form of the metric to Melvin's coordinates.
We also review the transverse space geometry of the wave front.

In Section \ref{sc:fieldeq}, we derive the field equations
and we simplify them introducing potentials. We discuss the structure of
the equations and the gauge freedom of the solutions.

Next, in Section \ref{sc:examples}, we solve the Einstein--Maxwell equations
in the special case of the $\phi$-independent spacetimes,
especially with a thin matter source localized on the axis of symmetry.

In Section \ref{sc:interpret} we concentrate on the interpretation of the
gyraton solutions. We discuss the properties of the scalar polynomial
invariants and the geometric properties of the principal null congruence.
We evaluate the curvature tensor in an appropriate tetrad, discuss the
Petrov type, the matter content of the spacetime, and properties of the electromagnetic field.

The main results of the paper are summarized in concluding Section \ref{sc:conclusion}.
Some technical results needed to derive the field equations,
spin coefficients and invariants are left to
Appendices \ref{apx:AppA}, \ref{apx:NP}, and \ref{apx:Invars}.

%%%%%%%%%%%%%%%%%%%%%%%%%%%%%%%%%%%%%%%%%%%%%%%%%%%%%%%%%%%%%%%%%%%%%%%%%%%%%%%%%%%%%%%%%%
\section{The gyratons on the Melvin spacetime}\label{sc:gyreq}
%%%%%%%%%%%%%%%%%%%%%%%%%%%%%%%%%%%%%%%%%%%%%%%%%%%%%%%%%%%%%%%%%%%%%%%%%%%%%%%%%%%%%%%%%

%%%%%%%%%%%%%%%%%%%%%%%%%%%%%%%%%%%%%%%%%%%%%%%%%%%%%%%%%%%%%%%%%%%%%%%%%%%%%%%%%%%%%%%%%
\subsection{The Melvin universe}\label{scc:melvin}
In this section we briefly review basic properties of the Melvin spacetime
\cite{Bonnor:1954:PRS:,Melvin:1965:PHYSR:,Ortaggio:2004:PHYSR4:} which
will be useful throughout the paper.
The Melvin universe describes an axial electromagnetic field concentrating under
influence of its self-gravity. The strength of the electromagnetic
field is determined by the parameters $E$ and $B$.
In cylindrical coordinates $(t,z,\rho,\phi)$, the metric and the Maxwell tensor read
\begin{gather}
d s^2=\Sigma^2(-\grad t^2+\grad z^2+\grad \rho^2)+\Sigma^{-2}\rho^2\grad \phi^2\;,\label{o1}\\
{F}=E\,\grad z \wedge \grad t + B\Sigma^{-2}\rho\,\grad \rho \wedge \grad \phi\;,\quad\label{Fintz}
\end{gather}
where
\begin{equation}\label{sigma}
\Sigma=1+\frac{1}{4}\vrho\rho^2\;.
\end{equation}
The constant $\vrho$ is given by the parameters $E$, $B$ as\footnote{%
$\varkappa=8\pi G$ and $\epso$ are gravitational and electromagnetic constants.
There are two standard choices of geometrical units: the Gaussian with $\varkappa=8\pi$ and
$\varepsilon_{\rm o}=1/4\pi$, and SI like with $\varkappa=\varepsilon_{\rm o}=1$.}
\begin{equation}\label{rhodef}
\vrho=\frac{\varkappa\epso}{2}(E^2+B^2)
%=\frac{\varkappa\epso}{2}\mid{\mathcal{B}}\mid^2
\;.
\end{equation}

Introducing double null coordinates,
\begin{equation}\label{null}
u=\frac{1}{\sqrt{2}}(t-z)\;,\quad v=\frac{1}{\sqrt{2}}(t+z)\;,
\end{equation}
we obtain an alternative expression for metric \eqref{o1}
and the Maxwell tensor \eqref{Fintz}:
\begin{gather}
d s^2=\Sigma^2(-2\grad u\,\grad v+\grad \rho^2)+\Sigma^{-2}\rho^2\grad \phi^2\;,\label{o2}\\
{F}=E\,\grad v \wedge \grad u + B\Sigma^{-2}\rho\,\grad \rho \wedge \grad \phi\;.\quad\label{physF}
\end{gather}
The electromagnetic field can be rewritten also in the complex self-dual form,\footnote{%
We follow the notation of \cite{Step:2003:Cam:}, namely, $\mathcal{F}\equiv {F}+i\,{{\star}F}$
is a complex self-dual Maxwell tensor, where the 4-dimensional Hodge dual is
${\star}F_{\mu\nu}=\frac{1}{2}\varepsilon_{\mu\nu\rho\sigma}F^{\rho\sigma}$.
The self-dual condition reads ${\star}\mathcal{F}=-i\mathcal{F}$.
The orientation of the 4-dimensional Levi-Civita tensor
is fixed by the sign of the component $\varepsilon_{vu\rho\phi}=\rho\Sigma^2$.
The energy-momentum tensor of the electromagnetic field is given by
$T_{\mu\nu}=\frac{\varepsilon_{\rm o}}{2}\mathcal{F}_{\mu}{}^{\rho}\overline{\mathcal{F}}_{\nu\rho}$.}
\begin{equation}
%{\mathcal{F}}=\mathcal{B}(\grad z \wedge \grad t - i\Sigma^{-2}\rho\,\grad \rho \wedge \grad \phi).
\mathcal{F}=\mathcal{B}\,\bigl(\grad v \wedge \grad u - i\Sigma^{-2}\rho\,\grad \rho \wedge \grad \phi\bigr)\label{mF}\;.
\end{equation}
with the complex constant $\mathcal{B}$ defined as
\begin{equation}\label{B}
\mathcal{B}=E+iB\;.
\end{equation}

The metric \eqref{o1} resembles a vacuum solution of the Levi-Civita family \cite{Step:2003:Cam:} for a large
value of $\rho$. For ${\vrho=0}$ (${E,B=0}$) spacetime reduces to the Minkowski spacetime in cylindrical coordinates.
For ${E\neq0}$, ${B=0}$ the Maxwell tensor describes
an electric field pointing along the $z$-direction,
whereas for ${E=0}$, ${B\neq0}$ we get a purely magnetic field oriented along the $z$-direction.

The metric admits the four Killing vectors
\begin{equation}
\partial_{t}\;,\quad \partial_{z}\;,\quad \partial_{\phi}\;,\quad z\partial_{t}+t\partial_{z}=v\partial_{v}-u\partial_{u}\;,
\end{equation}
which correspond to staticity, cylindrical symmetry, and invariance under a boost transformation.
Using the adapted null tetrad ${{k}=\partial_{v}}$, ${{l}=\Sigma^{-2}\partial_{u}}$, and
${{m}=\frac{1}{\sqrt{2}}(\Sigma^{-1}\partial_{\rho}-i\Sigma\rho^{-1}\partial_{\phi})}$,
the only non--vanishing components of Weyl and Ricci tensors are
\begin{equation}\begin{aligned}\label{Melpsi}
\Psi_{2}&=\frac{1}{2}\vrho\,(-1+\frac{1}{4}\vrho\rho^2)\Sigma^{-4}\;,\\
\Phi_{11}&=\frac{1}{2}\vrho\,\Sigma^{-4}\;.
\end{aligned}\end{equation}
This demonstrates that the Melvin universe is a non--vacuum solution
of the Petrov type D, except at points satisfying $\rho=2/\sqrt{\vrho}$, where the Weyl tensor vanishes.
It is interesting to note that the scalar curvature vanishes, $R=0$.

To conclude, the Melvin spacetime belongs to the family of non-expanding, non-twisting type D electrovacuum
solutions investigated by Pleba\'{n}ski \cite{Plebanski:1979:JMATHP:}. As a consequence, it also belongs
to the general Kundt class \cite{Step:2003:Cam:,GrifPod:2009:Cam:} which will be important in the following text.

%%%%%%%%%%%%%%%%%%%%%%%%%%%%%%%%%%%%%%%%%%%%%%%%%%%%%%%%%%%%%%%%%%%%%%%%%%%%%%%%%%%%%%%%%
\subsection{The ansatz for the gyratons on Melvin universe}\label{ssc:def}

Gyratons are generalized gravitational waves corresponding to null sources with intrinsic rotation.
In general, the gyraton solutions are obtained by adding non-diagonal terms to the metric of the standard
gravitational wave solutions, or in other words, by keeping the non-diagonal terms $g_{ui}$ in
the standard Kundt metric \cite{Step:2003:Cam:}.
Therefore, we derive the ansatz for the gyraton on Melvin spacetime by adding such new
terms to the Kundt form of the Melvin metric.
It can be explicitly obtained from \eqref{o2} by transformation
\begin{equation}\label{r-v}
v=\Sigma^{-2}r\;,
\end{equation}
which leads to
\begin{equation}\label{MK}
d s^2=-2\grad u\,\grad r+d s^2_{\trans}+2 r W_i\,\grad u\grad x^i\;.
\end{equation}
Here we introduced a 2-dimensional metric
\begin{equation}\label{transmM}
d s^2_{\trans}=\Sigma^2\grad \rho^2+\Sigma^{-2}\rho^2\grad \phi^2
\end{equation}
and ${r}$-independent 1-form $W=W_i\grad x^i$,
\begin{equation}
W=2\frac{\Sigma_{,\rho}}{\Sigma}\grad \rho\;.
\end{equation}
These tensors can be understood as tensors on space spanned by two coordinates ${\rho,\,\phi}$. This space
can be covered by other suitable spatial coordinates ${x^i}$, and we will use the Latin indices
${i,j,\dots}$ to label the corresponding tensor components.

By an appropriate transformation of coordinates
\cite{Ortaggio:2004:PHYSR4:,GrifPod:2009:Cam:}, the 2-dimensional metric $ds^2_{\trans}$
can be transformed into a conformally flat form,
which in the standard complex null coordinates $\zeta,\,\bar{\zeta}$ reads
\begin{equation}\label{trmetric}
d s^2_{\trans}= \frac{2}{P^2}\grad \zeta \grad \bar{\zeta}\;.
\end{equation}
%The corresponding components of the \mbox{1-form} ${W}$ are\\\textsf{???????????????}
%\begin{equation}\label{Wepr}
%W_{\zeta}=W_{\bar{\zeta}}=\frac{\sqrt{2}}{P^2\Sigma}\;.
%W_{\zeta}=W_{\bar{\zeta}}=\frac{\Sigma_\rho}{\sqrt{2}P\Sigma^2}\;.
%\end{equation}
%The details of the transformation between $\rho,\,\phi$ and $\zeta,\,\bar{\zeta}$ can be found in \cite{Ortaggio:2004:PHYSR4:,GrifPod:2009:Cam:}.
Such a transformation brings the metric \eqref{MK} into the Kundt form.

The gyraton generalization of \eqref{MK} then reads
\begin{equation}\begin{split}\label{WR}
d s^2&=-2\Sigma^2{H}\,\grad u^2 - 2 \grad u \grad r+d s^2_{\perp}\\
&\quad+ 2(r W_i {-} \Sigma^2 a_i)\; \grad u \grad x^i\;.
\end{split}\end{equation}
We have added the term ${-2\Sigma^2{H}\,\grad u^2}$ which represents a gravitational wave on the
Melvin universe \cite{GarfinkleMelvin:1992:PHYSR4:,Ortaggio:2004:PHYSR4:} with an arbitrary profile function ${H}$,
and the non-diagonal terms ${-2\Sigma^{2}a_i\grad u\grad x^i}$ characteristic for gyratons.
It will be shown in the following that these terms can be generated by specific gyratonic matter.
%The additional gyratonic terms ${a_i}$ are assumed to be $r$-independent, which would also follow from the Maxwell equations.

Transforming back to the Melvin coordinate $v$ and cylindrical coordinates ${\rho,\,\phi}$,
we obtain the ansatz for the metric describing the gyraton on Melvin spacetime,\footnote{%
Here we use notation different from \cite{GarfinkleMelvin:1992:PHYSR4:,Ortaggio:2004:PHYSR4:},
we use $-2H\grad u^2$ instead of $-H\grad u^2$ to match our notation in \cite{Kadlecova:2009:PHYSR4:}.}
\begin{equation}\begin{split}\label{s5}
d s^2&=-2\Sigma^2 H \grad u^2-2\Sigma^2\grad u\,\grad v  + \bigl(\Sigma^2\grad \rho^2 + \Sigma^{-2}\rho^2\grad \phi^2\bigr) \\
&\quad+2\,\Sigma^2 \bigl(a_{\rho}\,\grad u\,\grad \rho+ a_{\phi}\,\grad u\,\grad \phi\bigr)\;.
\end{split}\raisetag{13pt}\end{equation}
%An explicit relation to complex components ${a_\zeta,\,a_{\bar\zeta}}$ are
%\textsf{???????????????}
%\begin{equation}\begin{aligned}
%&a_{\rho}(u,\rho,\phi)=-\frac{2\sqrt{2}\,\Sigma^2}{\vrho^2\rho}(a_{\zeta}+a_{\bar{\zeta}})\;,\\
%&a_{\phi}(u,\rho,\phi)=-i\frac{2\sqrt{2}}{\vrho^2}(a_{\zeta}-a_{\bar{\zeta}})\;.
%\end{aligned}\end{equation}

The function $H(u,v,\rho,\phi)$ can depend on all coordinates, but we assume that the
functions ${a_{i}(u,\rho,\phi)}$ are ${\mbox{${v}$ independent}}$
(it actually follows from the Maxwell equations as will be shown below).
Let us note that the previously cited works assumed also the function $H$  \mbox{$v$ independent}.

%%%%%%%%%%%%%%%%%%%%%%%%%%%%%%%%%%%%%%%%%%%%%%%%%%%%%%%%%%%%%%%%%%%%%%%%%%%%%%%%%%%%%%%%%
\subsection{The ansatz for the matter}\label{ssc:ansatz}

The metric should satisfy the Einstein equations with a stress-energy tensor generated by
the electromagnetic field and the gyratonic source,
\begin{equation}\label{EinsteinEq}
G_{\mu\nu}=\varkappa \bigl( T^\EM_{\mu\nu}+T^{\gyr}_{\mu\nu}\bigr)\;.
\end{equation}
%Here, $\varkappa=8\pi G$ is the gravitational constant.
We assume that the electromagnetic field \eqref{physF} modified by the gyraton is given by
\begin{equation}\label{realF}
{F}=E\,\grad v \wedge\grad u+B\Sigma^{-2}{\rho}\,\grad \rho\wedge\grad \phi+\sreal_{j}\grad u\wedge\grad x^{j}\;.
\end{equation}
Similar to \cite{Kadlecova:2009:PHYSR4:}, we have added the term  $\sreal_{j}\,\grad u\wedge\grad x^{j}$.
%In fact, such new terms are generated if we demand a gauge symmetry of the electromagnetic field under gauge transformation discussed in \ref{ssc:gauge}.

To evaluate Maxwell equations, it is useful to write down the self-dual form of the Maxwell tensor $\mathcal{F}$.
The Hodge dual of \eqref{realF} reads
\begin{equation}\begin{split}\label{realF1}
{\star}{F}&=B\,\grad v\wedge\grad u-E\Sigma^{-2}{\rho}\,\grad \rho\wedge\grad \phi\\
&\qquad\qquad+({*\sreal}+Ba-E\,{*a})_{j}\,\grad u\wedge\grad x^{j}\;,
\end{split}\end{equation}
where the star $*$ means the 2-dimensional Hodge dual defined on the transversal space,
see the next section \eqref{ssc:transsp}. For the self-dual Maxwell tensor ${\mathcal{F}}$ we thus obtain
\begin{equation}\begin{split}\label{EMF}
\mathcal{F}&=\mathcal{B}\,\Bigl(\grad v\wedge\grad u - i\Sigma^{-2}\rho\,\grad \rho\wedge\grad \phi\\
 &\qquad\qquad\qquad+(\mathcal{S}-i\,{*a})_{j}\,\grad u\wedge\grad x^{j}\,\Bigr)\;.
\end{split}\end{equation}
where we introduced a complex transverse \mbox{1-form}~$\mathcal{S}_{j}$,
\begin{equation}
\mathcal{B}\mathcal{S}_{j}=(\sreal+i{*\sreal})_{j}+iB(a+i{*a})_{j}\;.
\end{equation}
This form is self-dual with respect to the Hodge duality ${*}$ on the transversal space,
\begin{equation}
*\mathcal{S}_{j}=-i\,{\mathcal{S}}_{j}\;,
%\mathcal{S}_{j}-i\,{*\mathcal{S}}_{j}=0\;,
\end{equation}
and therefore it can be written using a real \mbox{1-form}~$\sslfdl_{j}$:
\begin{equation}
\mathcal{S}_{j}=(\sslfdl+i\,{*\sslfdl})_{j}\;.
\end{equation}
%Then we can write the components in front of the term $\grad u\wedge\grad x^{j}$ as
%\begin{equation}
%(\mathcal{S}-i{*a})_{j}=(\sslfdl+i{*(\sslfdl-a)})_{j}.
%\end{equation}
The original 1-form $\sreal_{i}$ can be expressed in terms of $\sslfdl_{j}$ as
\begin{equation}\label{si}
\sreal_{j}=E\,\sslfdl_{j}-B\,{*(\sslfdl-a)_{j}}\;.
\end{equation}
In the following we use $\sslfdl_{j}(v,u,\rho,\phi)$ as a basic variable for the
electromagnetic field.

The stress-energy tensor ${T^\EM_{\mu\nu}}$ corresponding to the field \eqref{EMF} is given in \eqref{EMT}.

Finally, we must define the gyratonic matter by specifying the structure of its stress-energy tensor.
It is obtained from the standard stress-energy tensor of a null fluid by adding terms corresponding
to `internal spatial rotation' of the fluid:
\begin{equation}\label{m7}
\varkappa\, T^{\gyr}=j_{u}\,\grad u^2+2j_{\rho}\,\grad u\,\grad \rho+2j_{\phi}\,\grad u\,\grad \phi\;.
\end{equation}
We admit a general coordinate dependence of the source functions ${j_u(v,u,\rho,\phi)}$ and ${j_j(v,u,\rho,\phi)}$.
However, it will be shown below that the field equations enforce a trivial ${v}$ dependence.

The gyraton source is described only on a phenomenological level, by its
stress-energy tensor \eqref{m7}, which is assumed to be given, and our aim is to
determine its influence on the metric and the electromagnetic field.
However, we have to consider that the gyraton stress-energy tensor is locally conserved.
It means that the functions ${j_u}$ and ${j_i}$ must satisfy the constraint given by
\begin{equation}\label{gyrenergycons}
  T^{\gyr}_{\;\,\mu\nu}{}^{\>;\nu}=0\;.
\end{equation}
Of course, if we had considered a specific internal structure of the gyratonic matter,
the local energy-momentum conservation would have been a consequence of field equations
for the gyraton. Without that, we have to require \eqref{gyrenergycons} explicitly.

To conclude, the fields are characterized by functions ${\Sigma}$, ${H}$, ${a_j}$, and ${\sslfdl_j}$,
which must be determined by the field equations provided the gyraton sources ${j_u}$ and ${j_j}$
and the constants ${E}$ and ${B}$ are given.

%%%%%%%%%%%%%%%%%%%%%%%%%%%%%%%%%%%%%%%%%%%%%%%%%%%%%%%%%%%%%%%%%
\subsection{The geometry of the transverse space}\label{ssc:transsp}

The geometry \eqref{s5} identifies the null geodesic congruence
generated by $\partial_v$ which is parametrized by an affine parameter $v$,
the family of null hypersurfaces ${u=\text{constant}}$, and 2-dimensional
\textit{transverse spaces} ${u,v=\text{constant}}$.

The gravitational wave moves along the null direction ${\partial_v}$,
i.e., it propagates with the speed of light along the $z$-direction,
which is the direction of the electromagnetic field.
The hypersurface ${u=\text{constant}}$ corresponds to the surface of the constant `phase', and
the transverse spaces ${u,v=\text{constant}}$ are spatial wave fronts of the wave.

Physical quantities do not depend on the affine parameter ${v}$, or this dependence is trivial
and it will be explicitly found. Specifically, the geometry of the transverse space is \mbox{${v}$ independent}.

It turns out to be convenient to restrict various quantities to the transverse space.
For example, we can interpret ${a_i}$ and ${\sslfdl_i}$ as components of
$u$-dependent ${\mbox{${1}$-forms}}$ on the transverse space.
Our goal is to formulate all equations for physical quantities
on the transverse spaces. For that we need to review some properties of the
transverse geometry. It was studied in detail in \cite{Ortaggio:2004:PHYSR4:},
and on a general level in \cite{Kadlecova:2009:PHYSR4:}, nevertheless
it will be useful to mention some of the properties explicitly.

The transverse metric is obtained by restriction of the full metric \eqref{s5}
to the transverse space and it is given by the expression \eqref{transmM},
\begin{equation}\label{transm}
d s^2_{\trans}=g_\trans{}_{ij}\grad x^i \grad x^j = \Sigma^2\grad \rho^2+\Sigma^{-2}\rho^2\grad \phi^2
\end{equation}

The associated Gauss curvature is given by the scalar curvature $R_{\trans}$,
\begin{equation}\label{trsccurv}
K=\frac12 R_\trans=\frac{\vrho}{\Sigma^{4}}\Bigl(2-{\textstyle\frac{1}{4}}\vrho\rho^2\Bigr)\;.
\end{equation}
It is obvious that only the electromagnetic field is responsible for the
non-flatness of the transversal space---it is insensitive to the presence of the gyraton.
For $\vrho=0$ the curvature vanishes and we get the flat plane.

In general, the curvature is not constant and it is finite everywhere.
The Gauss curvature $K$ has maximum on the axis $\rho=0$ where it is equal to $2\vrho$;
it is positive for $0\leq\rho<2\sqrt{2}/\sqrt{\vrho}$, and vanishes on the circle at $\rho=2\sqrt{2}/\sqrt{\vrho}$.
For $\rho>2\sqrt{2}/\sqrt{\vrho}$, it goes to negative values and it has its minimum $K=-\frac{\vrho}{256}$ at $\rho=2\sqrt{3}/\sqrt{\vrho}$.
Then it grows again, and as $\rho\rightarrow +\infty$ the curvature vanishes, $K\rightarrow 0^{-}$.

The circumference of a circle of constant radius $\rho$ is vanishing when $\rho\rightarrow+\infty$.
Therefore, we measure a much shorter circumference for larger $\rho$---as if we would move
``along the stem of the wine-glass toward the narrowing end,''
\cite{Melvin:1965:PHYSR:,Thorne:1965:PHYSR:,Ortaggio:2004:PHYSR4:}.

The 2-dimensional Levi-Civita tensor ${\epsilon_{ij}}$ associated with the metric \eqref{transm}
is ${\epsilon=\rho\,\grad\rho\wedge\grad\phi}$. The covariant derivative will be denoted by
a colon, e.g., ${a_{i:j}}$. We raise and lower the Latin indices using ${g_\trans{}_{ij}}$,
which differs from lowering indices using $g_{\alpha\beta}$ thanks to non-vanishing terms $g_{ui}$.
We use a shorthand for a transverse square of the norm of a 1-form ${a_i}$ as
%\begin{equation}
%a^2\equiv a^i a_i=\frac{1}{\Sigma^2}a_\rho^2+\frac{\Sigma^2}{\rho^2}a_\phi^2\;.
%\end{equation}
\begin{equation}
a^2\equiv a^i a_i=\Sigma^{-2}a_\rho^2+\rho^{-2}\Sigma^2 a_\phi^2\;.
\end{equation}

In two dimensions, the Hodge duals of 0, 1 and 2-forms ${\varphi}$, ${a_i}$, and ${f_{ij}}$, respectively, read
\begin{equation}
(*\varphi)_{ij} = \varphi\, \epsilon_{ij}\;,\;\;
(*a)_i = a_j \epsilon^j{}_i\;,\;\;
*f = \frac12 f_{ij}\epsilon^{ij} =\frac{1}{\rho}f_{\rho\phi} \;.
\end{equation}

For convenience, we also introduce an explicit notation for 2-dimensional divergence and rotation of a transverse 1-form ${a_i}$,
\begin{align}
  &\div a \equiv a_i{}^{:i}= \frac{1}{\Sigma^2}a_{\rho,\rho}+\frac{\Sigma^2}{\rho^2}a_{\phi,\phi}+\frac{1}{\rho\Sigma^2}a_{\rho}-\frac{\Sigma^2_{,\rho}}{\Sigma^4}a_{\rho}\;,\notag\\
&\rot a \equiv \,\epsilon^{ij} a_{j,i} = \frac{1}{\rho}(a_{\phi,\rho}-a_{\rho,\phi})\;.
\end{align}
For 2-form $f_{ij}$ we get
\begin{equation}
  \div f \equiv f_{ij}{}^{:j} = \frac{1}{\Sigma^2}(f_{\phi\rho,\rho}-\frac{1}{\rho}f_{\phi\rho})\grad \phi+\frac{\Sigma^2}{\rho^2}f_{\rho\phi,\phi}\grad \rho\;,
\end{equation}
and ${\rot f=0}$. We can generalize the action of divergence and rotation also on a scalar function $f$ as
${\div f = 0}$ and ${\rot f = -{*}\grad f}$.
Note that the divergence and rotation are related as ${\div a = \rot {*}a}$, and
the relation to the transverse exterior derivative is ${\grad a=*\rot a}$.
Clearly, ${\div\div a=0}$, ${\div\rot a=0}$, and ${\rot \grad a=0}$.

The Laplace operator of a function ${\psi}$ reads
\begin{equation}\label{lapldef}
\laplace\psi =  \psi_{:i}{}^{:i} = \frac{1}{\Sigma^2}\psi_{,\rho\rho}+\frac{\Sigma^2}{\rho^2}\psi_{,\phi\phi}+\frac{1}{\rho\Sigma^2}\psi_{,\rho}-\frac{\Sigma^2_{,\rho}}{\Sigma^4}\psi_{,\rho}\;,
\end{equation}
and for a transverse 1-form~${\eta}$ it is defined as
${\laplace\eta \equiv \grad\,\div\eta-\rot\rot\eta}$.

Finally, the transverse space is topologically trivial since it has topology of a plane. We can thus assume that the
Poincare lemma (${\grad\omega=0} \Rightarrow {\omega=\grad\sigma}$) holds, which in terms of rotation and divergence means that ${\div\omega=0}$ implies ${\omega = \rot\sigma}$. However, since the transverse space is non-compact and we do not know a priori boundary conditions
for various quantities at infinity $\rho\to\infty$, we have to admit non-trivial harmonics. Therefore, we cannot assume
a uniqueness of the Hodge decomposition. Moreover, in some cases it can be physically relevant to consider also
topologically nontrivial harmonics which are singular, e.g., at the origin ${\rho=0}$. Such solutions would correspond
to fields around singular sources localized on the axis. However, we will ignore these cases in a general discussion.

%%%%%%%%%%%%%%%%%%%%%%%%%%%%%%%%%%%%%%%%%%%%%%%%%%%%%%%%%%%%%%%%%%%%%%%%%%%%%%%%%%%%%%%%%%
\section{The field equations}\label{sc:fieldeq}
%%%%%%%%%%%%%%%%%%%%%%%%%%%%%%%%%%%%%%%%%%%%%%%%%%%%%%%%%%%%%%%%%%%%%%%%%%%%%%%%%%%%%%%%%

%%%%%%%%%%%%%%%%%%%%%%%%%%%%%%%%%%%%%%%%%%%%%%%%%%%%%%%%%%%%%%%%%
\subsection{The field equations for matter}\label{scc:fequations}
%\subsection{The Maxwell and Einstein field equations}\label{scc:fequations}
Now, we will investigate the equations for matter, i.e., the Maxwell equations for electromagnetic field and the condition \eqref{gyrenergycons}
for the gyraton source.

Both Maxwell equations for real Maxwell tensor are equivalent to the cyclic Maxwell equation for the self-dual Maxwell tensor \eqref{EMF},
\begin{equation}\begin{aligned}\label{MXEC}
  \grad {\mathcal{F}}
   =\mathcal{B}&\Bigl[\,\partial_v\bigl(\sslfdl+i\,{*(\sslfdl-a)}\bigr)_{j}\,\grad v\wedge\grad u\wedge\grad x^j \\
   &-\bigl(\rot\sslfdl+i\,\div(\sslfdl-a)\bigr)\, \grad u \wedge \epsilon\,\Bigr]=0\;.
\end{aligned}\end{equation}
From the real part we immediately get that the 1-form~${\sslfdl_i}$ is ${v}$ independent, $\partial_v \sslfdl_i=0$, and rotation-free,
\begin{equation}\label{pot1}
  \rot\sslfdl=0\;.
\end{equation}
From the imaginary part  it follows that
the 1-form $a_{i}$ is also ${v}$ independent (as we have already mentioned above) and it satisfies
\begin{equation}\label{pot2}
  \div(\sslfdl- a) = 0\;.
\end{equation}

Equations \eqref{pot1} and \eqref{pot2} guarantee the existence and determine the structure of potentials
which will be discussed in detail in Section \ref{ssc:pots}.

Next, we analyze the condition \eqref{gyrenergycons} for the gyraton source. When translated to the transverse space, it gives
\begin{equation}\label{gyrenergycons2}
  -\partial_v j_i\,\grad x^i + \left(-\partial_v j_u+\div(\Sigma^2 j)+\Sigma^2 a^i \partial_v j_i\right)\grad u =0\;.
\end{equation}
The source functions ${j_i}$ must be thus ${v}$ independent and ${j_u}$ has to have the structure
\begin{equation}\label{jdecomp}
  j_u = v\,\div(\Sigma^2 j) + \iota\;,
\end{equation}
where $\iota(u,x^{i})$ is a $v$ independent function.
The gyraton source \eqref{m7} is therefore fully determined by
three \mbox{${v}$-independent} functions ${\iota(u,x^j)}$ and ${j_i(u,x^j)}$.

Equation \eqref{jdecomp} gives us also an insight into interpretation of the gyratonic terms
${j_i}$. They are composed from two contributions: one is related to a kind of `heat flow' which
changes energy ${j_u}$ of the fluid, and the other which is related to intrinsic rotation of the
fluid. The source representing `heat flow' has thus non-vanishing divergence ${\div(\Sigma^2 j)}$
and we require a vanishing rotational part ${\rot(\Sigma^2 j)}$. In opposite,
the source representing intrinsic rotation has vanishing heat flow, i.e., it satisfies
\begin{equation}\label{noheat}
  \div(\Sigma^2 j) =0\;.
\end{equation}
Such a source can be written in terms of a rotational potential ${\nu}$
as ${j=-\Sigma^{-2}\rot\nu}$. In components it means
\begin{equation}\label{noheatj}
   j_\rho = -\frac1\rho\nu_{,\phi}\;,\quad
   j_\phi = \frac\rho{\Sigma^4}\nu_{,\rho}\;.
\end{equation}

Physically more relevant is the rotational part of the source, since it can describe the spin of the
null fluid, or, in a specific limit, of the polarized beam of light. Terms related to heating flow
have bad causal behavior and therefore typically do not satisfy
various energy conditions and they are thus rather unphysical.
Interpretation of the gyraton source was discussed previously also in
\cite{Fro-Is-Zel:2005:PHYSR4:, Fro-Fur:2005:PHYSR4:,Fro-Zel:2005:PHYSR4:,Kadlecova:2009:PHYSR4:}.

%%%%%%%%%%%%%%%%%%%%%%%%%%%%%%%%%%%%%%%%%%%%%%%%%%%%%%%%%%%%%%%%%
\subsection{The Einstein equations}\label{scc:EinsteinEqs}

The Einstein gravitational law \eqref{EinsteinEq} needs the Einstein tensor and the electromagnetic and gyraton
stress-energy tensors. These quantities can be found in the Appendix \ref{apx:AppA}.
We can combine them and inspect various components of the Einstein equations.

The $vu$-component determines the function $\Sigma$, namely it gives the condition
\begin{equation}\label{Sig}
-\rho(\Sigma_{,\rho})^2+2\Sigma\Sigma_{,\rho}=\vrho\,\rho\;.
\end{equation}
It is straightforward to check that it is satisfied again by $\Sigma$ in the Melvin form \eqref{sigma}.

The transverse diagonal components $\rho\rho$ and $\phi\phi$ require
\begin{equation}\label{Einsteinii}
  \partial^2_{v} H  =0\;,
\end{equation}
thus we obtain the explicit ${v}$ dependence of the metric function ${H}$ as
\begin{equation}\label{Heq}
  H = g\,v + h\;,
\end{equation}
where we have introduced ${v}$-independent functions ${g(u,x^j)}$ and ${h(u,x^j)}$.

The remaining nontrivial components of the Einstein equations are those involving the gyraton source \eqref{m7}.
The $ui$-components give the equation related to ${j_i}$,
\begin{equation}\label{jieq}
  j_i= \frac12\, \Sigma^2 f_{ij}{}^{:j}  -(\Sigma^2)_{,k}\,g^{kj}f_{ji}+ g_{,i}+\frac{2\vrho}{\Sigma^2}(\sslfdl_{i}-a_{i})\;,
\end{equation}
where we introduce the exterior derivative ${f_{ij}}$ of the transverse \mbox{1-form} ${a_i}$ and its
Hodge dual ${b(u,x^j)}$,
\begin{gather}\label{fdef}
f_{ij} \equiv a_{j,i}-a_{i,j} = (*\,b)_{ij}\;,\\
b \equiv * f = \rot a\;.\label{bdef}
\end{gather}
In terms of ${b}$, equation \eqref{jieq} can be rewritten as
\begin{equation}\label{jieqpot}
  \Sigma^2\,j = \frac12\,\rot(\Sigma^4\, b) + \Sigma^2{\grad}g+2\vrho(\sslfdl-a)\;.
\end{equation}
%\begin{equation}\begin{aligned}\label{jieqpot}
%  \Sigma^2\,j &= \frac12\div(\Sigma^4 f) + \Sigma^2{\grad}g+2\vrho(\sslfdl-a),\;\\
%   &= \frac12\rot(\Sigma^4\, b) + \Sigma^2{\grad}g+2\vrho(\sslfdl-a)\;,
%\end{aligned}\end{equation}
Here and in the following $\grad g$ represents just the transverse gradient ${\grad g=g_{,i}\,\grad x^{i}}$,
and for simplicity we skipped the transverse indices.

It is useful to split the equation into divergence and rotation parts by applying ${\div}$ and ${\rot}$:
\begin{align}
  \div{(\Sigma^2\, j)}&=\div {(\Sigma^2{\grad}g)},\label{divjeq}\\
  \rot{(\Sigma^2 j)}&= - \frac12 \laplace(\Sigma^4 b) + \rot(\Sigma^2{\grad}g)-2\vrho\,b\;. \label{rotjeq}
\end{align}
Here we have used the relations \eqref{pot1} and \eqref{bdef}.
The formula \eqref{divjeq} is the equation for $g$, \eqref{rotjeq} is the equation
for $b$ and together with \eqref{bdef} it determines $a_{i}$. In the next section we will return
to these equations introducing suitable potentials which allow us to escape the necessity
of taking an additional derivative of \eqref{jieqpot} when deriving the equation for ${a_i}$.

Finally, from the $uu$-component of the Einstein equation we obtain
\begin{equation}
\begin{split}
  j_u =\,&\div {(\Sigma^2{\grad}g)}\;v
      +\Sigma^2(\laplace h - {(\Sigma^{-2})_{,\rho}}h_{,\rho})\\
      &+\frac12\Sigma^4 b^2+ 2\Sigma^2 a^i g_{,i}+\partial_u\div(\Sigma^2 a) + g\, \div(\Sigma^2 a)\\
      &-2\vrho\,(\sslfdl-a)^2\;.\label{jueq}
\end{split}\raisetag{14pt}
\end{equation}
When we compare the coefficient in front of ${v}$ with \eqref{divjeq} we find that it has structure consistent with \eqref{jdecomp}.
The nontrivial ${v}$-independent part of \eqref{jueq} gives the equation for the metric function~${h}$,
%\begin{equation}\label{heq}
%\begin{split}
% &\laplace h - (\Sigma^{-2})_{,\rho}\,h_{,\rho}=
%      \Sigma^{-2}\iota \, -\frac12\Sigma^2 b^2- 2 a^i g_{,i}\\
%      &\quad -\frac1{\Sigma^2}\Bigl(
%      \partial_u\div(\Sigma^2 a) + g\, \div(\Sigma^2 a)-2\vrho\,(\sslfdl{-}a)^2
%      \Bigr)\;.
%% \\
%% \Sigma^2&\bigl(\laplace h - (\Sigma^{-2})_{,\rho}\,h_{,\rho}\bigr)=
%%      \iota \, -\frac12\Sigma^4 b^2- 2\Sigma^2 a^i g_{,i}\\
%%      & -\partial_u\div(\Sigma^2 a) - g\, \div(\Sigma^2 a)+2\vrho\,(\sslfdl-a)^2\;.
%% \\
%% \Sigma^2&\bigl(\laplace h + \Sigma^{-4}(\Sigma^2)_{,\rho}\,h_{,\rho}\bigr)=
%%      \iota \, -\frac12\Sigma^4 b^2- 2\Sigma^2 a^i g_{,i}\\
%%      & -\partial_u\div(\Sigma^2 a) - g\, \div(\Sigma^2 a)+2\vrho\,(\sslfdl-a)^2\;.
%% \\
%% \Sigma^2&\bigl(\laplace h + 2\Sigma^{-3}\Sigma_{,\rho}\,h_{,\rho}\bigr)=
%%      \iota \, -\frac12\Sigma^4 b^2- 2\Sigma^2 a^i g_{,i}\\
%%      & -\partial_u\div(\Sigma^2 a) - g\, \div(\Sigma^2 a)+2\vrho\,(\sslfdl-a)^2\;.
%\end{split}\raisetag{40pt}
%\end{equation}
\begin{equation}\label{heq}
\begin{split}
\Sigma^2&\bigl(\laplace h - (\Sigma^{-2})_{,\rho}\,h_{,\rho}\bigr)=
      \iota \, -\frac12\Sigma^4 b^2- 2\Sigma^2 a^i g_{,i}\\
      & -\partial_u\div(\Sigma^2 a) - g\, \div(\Sigma^2 a)+2\vrho\,(\sslfdl-a)^2\;.
\end{split}\raisetag{40pt}
\end{equation}

%%%%%%%%%%%%%%%%%%%%%%%%%%%%%%%%%%%%%%%%%%%%%%%%%%%%%%%%%%%%%%%%%
\subsection{Introducing potentials}\label{ssc:pots}

In the previous section we have found that the Maxwell and Einstein equations
reduce to two potential equations \eqref{pot1}, \eqref{pot2},
and two source equations \eqref{jieq}, \eqref{jueq}.

According to the two dimensional Hodge decomposition we can express the 1-form ${a_i}$
using two scalar potentials ${\kappa(u,x^j)}$ and ${\lambda(u,x^j)}$,
\begin{equation}\label{klpotdef}
    %a_i = \kappa_{,i}+{\epsilon_{i}}^{j}\lambda_{,j}\;.
    a = \grad\kappa+\rot\lambda\;.
\end{equation}
These potentials control the divergence and the rotation of~${a_i}$ as
\begin{equation}\label{divrota}
    \div a = \laplace\kappa\;,\quad \rot a = -\laplace\lambda\;.
\end{equation}
Comparing with \eqref{bdef} we thus obtain the equation for $\lambda$ in terms of $b$,
\begin{equation}
  \laplace\lambda = -b\;.\label{lambdab}
\end{equation}

The first potential equation \eqref{pot1} gives immediately that ${\sslfdl_i}$ has a potential ${\varphi(u,x^j)}$,
\begin{equation}\label{phipot}
  \sslfdl = {\grad}\varphi\;.
\end{equation}
Equation \eqref{pot2} implies that there exists a potential  ${\psi(u,x^j)}$ satisfying
\begin{equation}\label{potpsi}
   \sslfdl-a=-\rot\psi\;.
\end{equation}
In terms of these potentials the 1-form \eqref{si} from the real Maxwell tensor \eqref{physF} reads
\begin{equation}\label{sreal}
   \sreal=E\grad\varphi + B\grad\psi\;.
\end{equation}

The potential ${\varphi}$ and ${\psi}$ are not, however, independent.
Substituting \eqref{klpotdef} and \eqref{phipot} into \eqref{potpsi}
we obtain the key relation among the potentials:
\begin{equation}\label{potrel}
   {\grad}(\varphi -\kappa)+\rot(\psi-\lambda)=0\;.
\end{equation}
If the Hodge decomposition was unique, the gradient and rotational parts would be vanishing separately,
i.e., we would get ${\varphi = \kappa}$ and ${\psi = \lambda}$ (up to unimportant constants).
The non-uniqueness of the Hodge decomposition is linked to
the possible existence of a non-trivial harmonic 1-form $\eta_i(u,x^{j})$,
\begin{equation}\label{eta}
   \laplace\eta=0\;,
\end{equation}
in terms of which the gradient and rotation parts of \eqref{potrel} can be expressed as
\begin{align}
   \varphi-\kappa&=\div\eta\;,\label{phikappa}\\
   \psi-\lambda&=-\rot\eta\;.\label{psilambda}
\end{align}

These are equations for electromagnetic potentials ${\varphi}$ (or ${\psi}$, respectively)
in terms of the metric potentials ${\kappa}$ and ${\lambda}$. The 1-form ${\eta}$ encodes an extra freedom,
which allows a nontrivial electromagnetic field not uniquely determined by the metric. (Such contributions
would allow one to take into account, for example, an additional electromagnetic charge localized at the origin
of the transverse space,
cf.\ the discussion of particular cases in section \ref{sc:examples}.) However, sufficiently restrictive
conditions at the infinity and the smoothness on the whole transverse space for the potentials would
eliminate this freedom, so the case ${\eta=0}$ is a rather representative choice.

After eliminating the electromagnetic potentials, we need
to formulate the equations for ${\kappa}$ and ${\lambda}$. %and for the metric functions ${g}$, ${h}$.
We start with the divergence part of \eqref{jieqpot} which can be rewritten using
the modified Laplace operator acting on the metric function $g$:
\begin{equation}\label{geq}
\Sigma^2\bigl(\laplace g - (\Sigma^{-2})_{,\rho}\, g_{,\rho}\bigr)=\div\bigl(\Sigma^2 j\bigr)\;.
%\laplace g - (\Sigma^{-2})_{,\rho}\, g_{,\rho}=\Sigma^{-2}\div\bigl(\Sigma^2 j\bigr)\;.
%\laplace g + 2\Sigma^{-3}\Sigma_{,\rho}\, g_{,\rho}=\Sigma^{-2}\div\bigl(\Sigma^2 j\bigr)\;.
%\laplace g + \Sigma^{-2}(\Sigma^2)_{,i}g^{ij}g_{,j}=\Sigma^{-2}\div\bigl(\Sigma^2 j\bigr)\;.
\end{equation}
However, for ${g}$ solving this equation, \eqref{divjeq} is also
the integrability condition for the quantity ${\Sigma^2(\grad g-j)}$.
It can thus be written in terms of a
potential ${\omega}$,
\begin{equation}\label{eq2}
\Sigma^2({\grad}g-j)=\rot{\omega}\;,
\end{equation}
or, more explicitly,
\begin{equation}\label{eq3}
{\grad}\omega=-\Sigma^2(\rot g+{*}j)\;.
\end{equation}

For the source \eqref{noheatj} without intrinsic `heating', the right-hand side of \eqref{geq}
is zero and the function ${\omega}$ has an additive contribution from the rotational potential
${\nu}$, namely it has to satisfy
\begin{equation}\label{omeganoheat}
{\grad}(\omega-\nu)=-\Sigma^2\rot g\;.
\end{equation}

The function $\omega$ contains information from the source $j$ and from the metric function $g$
relevant for the rotational part of equation \eqref{jieqpot}.
Indeed, substituting the potentials and ${\rot\omega}$
into \eqref{jieqpot} we obtain
\begin{equation}\label{rotjeqq1}
\rot\Bigl(\frac12 \Sigma^4 b-2\vrho\psi+\omega\Bigr)=0\;.
\end{equation}
Substituting \eqref{lambdab}, \eqref{psilambda}, and integrating
(absorbing an integration constant to ${\omega}$),
we derive the equation for the potential ${\lambda}$
\begin{equation}\label{lambdaeq}
\frac12 \Sigma^4\laplace\lambda+2\vrho\lambda=\omega + 2\vrho\,\rot\eta\;.
\end{equation}
Taking into account relations \eqref{psilambda} and \eqref{eta}, we obtain the alternative
equation for ${\psi}$
\begin{equation}\label{psieq}
\frac12 \Sigma^4\laplace\psi+2\vrho\psi=\omega \;.
\end{equation}

The other metric potential ${\kappa}$ remains unrestricted. This non-uniqueness is related to
the gauge freedom discussed in detail in section \ref{ssc:gauge}.
This coordinate freedom allows us to set the potential ${\kappa}$ to an arbitrary convenient form,
e.g., to eliminate it completely.

%%%%%%%%%%%%%%%%%%%%%%%%%%%%%%%%%%%%%%%%%%%%%%%%%%%%%%%%%%%%%%%%%
\subsection{Discussion of the field equations}\label{ssc:fieldeqssum}

We have thus formulated all field equations as equations on the transverse space. They are written
in a separated form, i.e., they can be solved one after the other:
First, one has to find harmonic 1-form ${\eta}$ satisfying equation~\eqref{eta}
and metric function ${g}$ satisfying \eqref{divjeq}. It allows one to integrate the function ${\omega}$
which together with ${\rot\eta}$ appears as a source in equation \eqref{lambdaeq} for the potential
${\lambda}$. Using the gauge freedom one can choose the other potential ${\kappa}$. Equations
\eqref{phikappa} and \eqref{psilambda} determine electromagnetic potentials and through equation
\eqref{si} the electromagnetic field. Finally, the remaining metric function ${h}$ is determined by
equation \eqref{heq}, in which the previously computed quantities contribute to the source on the right-hand side.

The most complicated field equations---\eqref{geq}, \eqref{heq}, and \eqref{lambdaeq}---are partial
differential equations on the \mbox{2-dimensional} space, which are solvable, at least in principle. They all contain
a modified Laplace operator, equations \eqref{geq} and \eqref{heq} for ${g}$ and ${h}$ the same one, namely
\begin{equation}\label{modlaplop}
  \Sigma^2\bigl(\laplace f - (\Sigma^{-2})_{,\rho}\, f_{,\rho}\bigr)\equiv\div\bigl(\Sigma^2 \grad f\bigr)\;.
\end{equation}
Solutions for particular cases (assuming rotational symmetry) will be discussed in section~\ref{sc:examples}.

It is important to observe that, except equation \eqref{heq} for ${h}$,
the field equations are linear. We can thus superpose two solutions
simply by adding the fields together. Only in the last step, when computing the source
for equation \eqref{heq}, one has to include total superposition of the fields
${g}$, ${a_i}$, and ${\sreal_i}$ since the expression for the source is non-linear.

Finally, we have not paid much attention to the
\mbox{$u$ dependence} of the studied quantities.
All metric functions, matter fields and sources can depend on the coordinate $u$
and this dependence does not enter the field equations except in one term on the right-hand side of equation \eqref{heq}.
The profile of the gyraton in the ${u}$ direction can thus be specified arbitrarily.
It corresponds to the fact that both matter and gravitational field move with the speed of light
and information on different hypersurfaces $u=\text{constant}$ evolves rather
independently.

Also the dependence of the metric and fields on the coordinate $v$ is very simple
and it was found for all quantities explicitly.

%%%%%%%%%%%%%%%%%%%%%%%%%%%%%%%%%%%%%%%%%%%%%%%%%%%%%%%%%%%%%%%%%
\subsection{The gauge transformation}\label{ssc:gauge}

The coordinate transformation ${\tilde v\to v = \tilde v-\chi(u,x^j)}$ accompanied
by the following redefinition of the metric functions and matter fields:
\begin{equation}\begin{gathered}\label{gauge}
v=\tilde v-\chi\;,\\
g=\tilde g\;,\quad
h=\tilde h+\tilde g\,\chi + \partial_u\chi\;,\quad
a_i=\tilde a_i-\chi_{,i}\;,\\
\sslfdl_{i}=\tilde{\sslfdl}_{i}-\chi_{,i}\;,\quad
\sreal_i=\tilde{\sreal}_i-E\,\chi_{,i}\;,\\
\kappa=\tilde \kappa-\chi\;,\quad \lambda=\tilde\lambda\;,\quad
\varphi=\tilde\varphi-\chi\;,\quad \psi=\tilde{\psi}\;,\\
\omega=\tilde{\omega}\;, \quad \eta=\tilde{\eta}\;,\\
j_i=\tilde j_i\;,\quad \iota=\tilde\iota+\chi\, \div j\;,
\end{gathered}\end{equation}
leaves the metric, the Maxwell tensor, and the gyraton stress-energy
tensor in the same form. Therefore, all the field equations remain the same.
This transformation is thus a pure gauge transformation and we can use it
to simplify the solution of the equations.

There are two natural choices of gauge: we can eliminate either
the metric potential ${\kappa}$ or the electromagnetic potential ${\varphi}$.
In the first case ${a=\rot\lambda}$ and ${\varphi=\div\eta}$.
In the latter case ${\sslfdl=0}$, ${\sreal=B\grad\psi}$, ${\kappa=-\div\eta}$, and
${a=\rot\psi}$.

Let us mention that in  \cite{Kadlecova:2009:PHYSR4:} an analogous gauge transformation
allowed us to choose also the metric function ${g}$. For the gyraton on the Melvin universe the metric
function ${g}$ decouples from ${\kappa}$ and it is gauge independent.

%In \cite{Kadlecova:2009:PHYSR4:}, the deficiency of equations occured, i.e. there was only one equation for $\kappa$ and $g$, and we were able to use the gauge freedom to simplify the Einstein equations and discuss several choices of gauges (e.g. set $g=0$). Here, we obtained additional equation for the function $g$ which is given by the source $j_{i}$. As we observed, the equation is integrability condition for $\omega$. We can use this remaining gauge freedom in $\varphi$ to specify the function $\chi=-\div\eta$ then $\varphi=\tilde{\varphi}$. Then the remaining gauge freedom is connected to un-uniqueness 1-form $\eta$.
%Or we can put locally the gyraton functions to zero $a_{i}=0$ while $(\chi_{,i}=-\tilde{a}_{i})$, but we can not make $a_{i}$ to be zero globally because the contour integral $\oint a_{i}(u,x^{i}){\grad}x^{i}$ around the position of the gyraton is generally non-zero.

The discussed gauge transformation has a clear geometrical meaning: it corresponds to a shift of the
origin of the affine parameter ${v}$ of the null congruence ${\partial_v}$. Note that such a
change redefines transverse spaces.
%For more detailed discussion and other possible gauge transformations see \cite{KrtousPodolsky} and \cite{Krtousetal}.

%Let us mention the remaining gauge freedom. The metric \eqref{s5}, the electromagnetic field \eqref{EMF}, and the gyraton stress-energy tensor \eqref{m7} keep the same form under a general reparametrization of the ${u}$-coordinate, ${\tilde u\to u=f(\tilde u)}$, accompanied by the rescaling ${\tilde v\to v=\tilde v/f'(\tilde u)}$ of  the ${v}$-coordinate, see \cite{Kadlecova:2009:PHYSR4:}.

%%%%%%%%%%%%%%%%%%%%%%%%%%%%%%%%%%%%%%%%%%%%%%%%%%%%%%%%%%%%%%%%%%%%%%%%%%%%%%%%%%%%%%%%%%%
\section{Special cases}\label{sc:examples}
%%%%%%%%%%%%%%%%%%%%%%%%%%%%%%%%%%%%%%%%%%%%%%%%%%%%%%%%%%%%%%%%%%%%%%%%%%%%%%%%%%%%%%%%%%%

In this section we will study the special solutions of the field equations. Namely, we restrict
to the axially symmetric situation, i.e., to the case when the geometry and the fields are invariant
under action generated by the rotational vector ${\partial_\phi}$. Further, we concentrate on
the gyraton generated by a thin beam of matter concentrated at the origin of the transverse
space which means on the axis of symmetry.

Thank to linearity mentioned at the end of section \ref{ssc:fieldeqssum},
we can discuss various special cases separately. However,
the geometry of the spacetime with gyraton is not merely a superposition of the individual contribution
since the nonlinear coupling in the metric function ${h}$.

We do not discuss in detail the ${u}$ dependence of the fields.
It does not enter the field equations, except in the source term
of equation \eqref{heq} for ${h}$. The \mbox{${u}$ dependence} of the gyraton
sources and corresponding dependence of other fields can thus be chosen arbitrarily.

%The reason that the ${u}$ profile of the gyraton can be arbitrary is
%that the matter, electromagnetic field and gravitation move with
%speed of light and information traveling on different hypersurfaces
%${u=\text{constant}}$ is relatively independent.

The symmetry assumption enforces that quantities
${a_i}$, ${g}$, ${h}$, ${\sreal_i}$, ${j_i}$, and ${\iota}$
are ${\phi}$-independent. It can induce a slightly weaker
condition on the potentials: typically, they have
a linear dependence on ${\phi}$.

The thin beam approximation requires that gyratonic matter
is concentrated at the origin ${o}$ of the transverse space given by ${\rho=0}$, i.e.,
${j_i}$ and ${\iota}$ should be distributions with the support at the origin.
However, we relax this condition slightly in the case  of
gyratonic `heat flow' discussed in section \ref{ssc:heating}.

In all discussed cases we use a natural gauge
\begin{equation}\label{gaugekappa0}
    \kappa = 0\;,\quad\text{i.e.,}\quad
    a = \rot \lambda\;.
\end{equation}
Together with the symmetry assumptions it implies ${\lambda_{,\phi}=\text{constant}}$.

\subsection{Pure gravitational gyraton}
\label{ssc:gravgyr}

We start with the simplest vacuum case: we set ${j_i=0}$ and ${\iota=0}$
and we assume no pure electromagnetic contribution, i.e., ${\eta=0}$.
The equation for ${g}$ has only a trivial solution ${g=g_\ro=\text{constant}}$,
the function ${\omega}$ is also a trivial constant and we obtain equation
\eqref{lambdaeq} with vanishing right-hand side. Taking into account that
${\lambda_{,\phi}=\text{constant}}$ we obtain that ${\lambda}$ must be
${\phi}$ independent (a ${\phi}$-linear term in ${\lambda}$ would require
an analogous term in the source) and we obtain an ordinary differential equation
\begin{equation}\label{lambdaeqrho}
    \frac1\rho\Bigl(\frac\rho{\Sigma^2}\lambda_{,\rho}\Bigr)_{,\rho}+\frac{4\vrho}{\Sigma^4}\lambda = 0\;.
\end{equation}
Substituting \eqref{sigma} for ${\Sigma}$, it is possible to
obtain two independent solutions, one regular at the origin,
%\begin{equation}\label{lambdagravgyr}
%    \lambda=-\gamma\frac{1-\frac34\vrho\rho^2}{1+\frac14\vrho\rho^2}\;,
%\end{equation}
\begin{equation}\label{lambdagravgyr}
    \lambda=-\gamma\Sigma^{-1}\Bigl(1-{\textstyle\frac34}\vrho\rho^2\Bigr)\;,
\end{equation}
and the other behaving as ${\log\rho}$ near the origin, which corresponds to
a delta source at the origin and it will be discussed in section \ref{ssc:spin}.

The metric 1-form ${a=\rot\lambda}$ has thus components
\begin{equation}\label{agravgyr}
    a_\rho = 0\;,\quad a_\phi= -\frac\rho{\Sigma^2}\lambda_{,\rho}=-\gamma\frac{2\vrho}{\Sigma^4}\rho^2\;,
\end{equation}
and its `strength' ${b=\rot a}$ is then
\begin{equation}\label{bgravgyr}
    b = -\gamma \frac{4\vrho}{\Sigma^5}\Bigl(1-{\textstyle\frac34}\vrho\rho^2\Bigr)\;.
\end{equation}
These can be plugged into \eqref{heq} which turns out to be
\begin{equation}\label{hsourcegravgyr}
\frac1\rho\bigl(\rho h_{,\rho}\bigr){}_{,\rho}=-\gamma^2\frac{8\vrho^2}{\Sigma^6}
   \Bigl(1-{\textstyle\frac14}\vrho\rho^2\Bigr)\Bigl(1-{\textstyle\frac94}\vrho\rho^2\Bigr)\;.
\end{equation}
It can be integrated explicitly; however, the result is rather long, so we skip it.

\subsection{Non-spinning beam with `heat' flow}
\label{ssc:heating}

As we discussed in section \ref{scc:fequations}, the gyraton source can have two contributions:
one, which changes gyratonic energy density ${j_u}$ and the other corresponding to
intrinsic rotation. Let us investigate the case when the gyratonic energy is concentrated
at the origin (a thin beam) but there is axially symmetric energy flow in the ${\partial_\rho}$
direction which accumulates energy at the beam (a `heating' process).
Namely, we assume that ${\partial_v j_u = \div(\Sigma^2 j)}$ is nonzero
only at the origin, so elsewhere the transverse flow ${j_i}$ must satisfy
 equation \eqref{noheat}. Such ${j_i}$ has the form
\begin{equation}\label{jheating}
    j_\rho = \frac{\alpha}{2\pi}\frac1\rho\;,\quad j_\phi = 0\;,
\end{equation}
and the increasing energy of the gyraton is given by
\begin{equation}\label{juheating}
    j_u = \alpha\, v\, \delta_o\;.
\end{equation}
Here, ${\delta_o}$ stands for the transverse delta function localized
at the origin ${o}$, normalized to the standard metric volume element
on the transverse space.

Since the `heating' is localized only at the origin,
the gyraton source \eqref{jheating} can be locally written using
the source potential ${\nu}$, cf. \eqref{noheatj}, as
\begin{equation}\label{nuheating}
    \nu = -\frac\alpha{2\pi}\phi\;.
\end{equation}
Note however, that the potential cannot be defined globally and it is not
well behaved at the origin.

We again assume no pure electromagnetic contribution, i.e., ${\eta=0}$.
The requirement of the axial symmetry enforces that the difference ${\Sigma^{-2}\rot\omega}$
between ${\grad g}$ and ${j}$, cf.~\eqref{eq2}, must be zero, i.e., ${\omega=0}$.
The ${\phi}$-independent metric function ${g}$ must thus satisfy ${g_{,\rho} = j_\rho}$ which gives
\begin{equation}\label{gheating}
    g = \frac\alpha{2\pi}\log\rho+g_\ro\;.
\end{equation}

The equation for ${\lambda}$ has again the form \eqref{lambdaeqrho} and in this case
we choose the trivial solution ${\lambda=0}$. With the gauge ${\kappa=0}$ we thus obtain
the only nontrivial field to be the metric function ${g}$, otherwise ${a_i=0}$ and ${\sigma_i=0}$.
The source for equation \eqref{heq} is also trivial, given only by ${\iota}$,
and we will study it in the next case.

\subsection{Non-spinning light beam}
\label{ssc:nospin}

A particular example of the gyraton source is standard null fluid. The thin non-spinning beam
localized at the origin is described by the source
\begin{equation}\label{jjunospin}
    \iota = \varepsilon \delta_o\;,\quad j_i =0\;.
\end{equation}
We can set all the fields except ${h}$ to be zero: ${g=0}$, ${a_i=0}$, and ${\sigma_i=0}$.
The equation for ${h}$ outside the origin is ${\rho^{-1}(\rho h_{,\rho}){}_{,\rho}= 0}$
%\begin{equation}\label{heqrho}
%    \frac1\rho\bigl(\rho h_{,\rho}\bigr){}_{,\rho}= 0\;,
%\end{equation}
which (with proper fixing of the source constant) gives
\begin{equation}\label{hnospin}
    h = \frac\varepsilon{2\pi}\log\rho\;.
\end{equation}

\subsection{Thin gyraton---spinning light beam}
\label{ssc:spin}

Finally we proceed with the most characteristic representant of the gyratonic matter.
It is a simple null beam of energy localized at the origin with no heating, which, however,
contains an intrinsic energy rotation. Since we have a point-like source at the transverse space,
we can speak about inner spin instead of a global rotational energy flow.

The gyraton source has a form\footnote{%
The exact structure of the singular source ${j_i}$ at the origin
can be read out from the singular solution \eqref{lambdaspin} of equation \eqref{lambdaeq} for
${\lambda}$ below.}
\begin{gather}
    \iota = \varepsilon \delta_o\;,\label{iotaspin}\\
    j= -\Sigma^{-2}\rot\nu\;,\quad\text{with}\quad \nu = \beta\delta_o\;.\label{jspin}
\end{gather}
The symmetry assumptions together with the no-heating requirement implies  ${g=\text{constant}}$, which
we choose to be zero in this case. Equation \eqref{eq2} also implies that ${\omega=\nu}$.
Ignoring again the electromagnetic contribution, ${\eta=0}$, we obtain the equation for ${\lambda}$
\begin{equation}\label{lambdaeqrhonu}
    \frac1\rho\Bigl(\frac\rho{\Sigma^2}\lambda_{,\rho}\Bigr)_{,\rho}+\frac{4\vrho}{\Sigma^4}\lambda = \Sigma^{-4}\nu\;.
\end{equation}
The solution of the homogeneous equation with a singular behavior ${\sim\log\rho}$
corresponding to the delta function \eqref{jspin} at the origin reads
\begin{equation}\begin{split}\label{lambdaspin}
    \lambda=-\frac{\beta}{2\pi}\Sigma^{-1}&\Bigl(1+{\textstyle\frac12}\vrho\rho^2-{\textstyle\frac3{64}}\vrho^2\rho^4-{\textstyle\frac1{768}}\vrho^3\rho^4\\
        &+{\textstyle\frac12}\bigl(1{-}{\textstyle\frac34}\vrho\rho^2\bigr)\log\bigl({\textstyle\frac14}\vrho\rho^2\bigr)\Bigr)\;.
\end{split}\raisetag{18pt}
\end{equation}
The multiplicative constant in the argument of the logarithm can be chosen arbitrary since it generates only
an additional homogeneous contribution of the form \eqref{lambdagravgyr}.

The metric 1-form ${a_i}$ and its rotation ${b}$ are
\begin{equation}\label{aspin}
\begin{split}
    a_\rho &=0 \;,\\
    a_\phi &= -\frac\beta{2\pi}\Sigma^{-4}\Bigl(
        1-{\textstyle\frac38}\vrho^2\rho^4-{\textstyle\frac1{32}}\vrho^3\rho^6-{\textstyle\frac1{768}}\vrho^4\rho^8\\
        &\qquad\qquad\quad-\vrho\rho^2\log\bigl({\textstyle\frac14}\vrho\rho^2\bigr)\Bigr)\;,
\end{split}\raisetag{18pt}
\end{equation}
and
\begin{equation}\label{bspin}
\begin{split}
    b &= 2\frac{\beta}{2\pi}\vrho\Sigma^{-5}\Bigl(
         \bigl(1{-}{\textstyle\frac34}\vrho\rho^2\bigr)\log\bigl({\textstyle\frac14}\vrho\rho^2\bigr)\\
         &\qquad+2+\vrho\rho^2-{\textstyle\frac3{32}}\vrho^2\rho^4-{\textstyle\frac1{384}}\vrho^3\rho^6\Bigr)\;.
\end{split}
\end{equation}
The source for equation \eqref{heq} becomes cumbersome and lengthy, but treatable, in principle.

\subsection{Electromagnetic wave}
\label{ssc:EMw}

In the previous examples we have ignored the possibility of a nontrivial electromagnetic field.
Namely, we have assumed that the electromagnetic field is given by the metric potentials
via relations ${\varphi=\kappa}$ and ${\psi=\lambda}$. However, we have already observed that
equation \eqref{potrel} admits also other solutions which we have parametrized using
a harmonic 1-form ${\eta}$. To include such solutions we should classify all 1-form harmonics
on the transverse space. But if we restrict to the axially symmetric fields we can solve
the potential equation \eqref{potrel} directly, without referring to ${\eta}$ explicitly.

Let us study a pure electromagnetic contribution to the matter, i.e., we assume ${\iota=0}$, ${j_i=0}$ here.
We can thus take a trivial vanishing solution for ${g}$ which implies ${\omega=0}$.

%Let us introduce abbreviations ${\tilde\varphi=\varphi-\kappa}$ and ${\tilde\psi=\psi-\lambda}$.
The symmetry assumptions tell us that 1-forms ${\sslfdl}$, ${\sreal}$, and ${a}$ are ${\phi}$ independent,
which implies that all potentials including ${\varphi-\kappa}$ and ${\psi-\lambda}$ can
be at most linear in ${\phi}$:
\begin{equation}\label{tildephipsi}
    \varphi-\kappa = \hat\varphi(\rho)+\varphi_\phi\,\phi\;,\quad
    \psi-\lambda = \hat\psi(\rho)+\psi_\phi\,\phi\;,
\end{equation}
${\varphi_\phi}$, ${\psi_\phi}$ being constants and ${\hat\varphi}$ and ${\hat\psi}$ functions of $\rho$ only.
Substituting into \eqref{potrel} we get
\begin{equation}\label{potrelsym}
    \Bigl(\hat\varphi_{,\rho}-\frac{\Sigma^2}{\rho}\psi_\phi\Bigr)\grad\rho+
    \Bigl(\varphi_\phi+\frac\rho{\Sigma^2}\hat\psi_{,\rho}\Bigr)\grad\phi=0\;,
\end{equation}
which implies
\begin{equation}\label{potrelsymcom}
    \hat\varphi_{,\rho}=\frac{\Sigma^2}{\rho}\psi_\phi\;,\quad
    \hat\psi_{,\rho}=-\frac{\Sigma^2}{\rho}\varphi_\phi\;.
\end{equation}
Taking into account \eqref{sigma}, it can be easily integrated and substituting
back to \eqref{tildephipsi} we obtain
\begin{equation}\label{tildephipsii}
\begin{aligned}
    \varphi &= \kappa+\psi_\phi\bigl(\log\rho+{\textstyle\frac18}\vrho\rho^2\bigr)+\varphi_\phi\,\phi\;,\\
    \psi &= \lambda-\varphi_\phi\bigl(\log\rho+{\textstyle\frac18}\vrho\rho^2\bigr)+\psi_\phi\,\phi\;.
\end{aligned}
\end{equation}

At this moment it is easier to solve equation \eqref{psieq} for ${\psi}$ with vanishing
right-hand side instead of equation \eqref{lambdaeq} for ${\lambda}$. It has solutions in the form
\eqref{lambdagravgyr} and \eqref{lambdaspin}. The metric potential ${\lambda}$ is then given
by the second of the equations \eqref{tildephipsi}. The metric potential ${\kappa}$ is vanishing thanks
to our gauge.

After choosing the solution for ${\psi}$, we can thus compute
all quantities ${a}$, ${b}$, ${\sslfdl}$, and ${\sreal}$
and substitute them to equation \eqref{heq} for ${h}$.

Let us mention that solution \eqref{tildephipsi} is singular at the origin.
A careful distributional calculation would show that equations for the potentials
\eqref{phipot} and \eqref{potpsi} may not be satisfied at the origin.
Tracing this singular term back to Maxwell equations, it could lead to non-vanishing
electric charges localized at the origin. However, since we have not
written down the Maxwell equations with sources, we do not discuss these
terms in more detail.

%%%%%%%%%%%%%%%%%%%%%%%%%%%%%%%%%%%%%%%%%%%%%%%%%%%%%%%%%%%%%%%%%%%%%%%%%%%%%%%%%%%%%%%%%%%
\section{Properties of the gyraton spacetimes}\label{sc:interpret}
%%%%%%%%%%%%%%%%%%%%%%%%%%%%%%%%%%%%%%%%%%%%%%%%%%%%%%%%%%%%%%%%%%%%%%%%%%%%%%%%%%%%%%%%%%%

%%%%%%%%%%%%%%%%%%%%%%%%%%%%%%%%%%%%%%%%%%%%%%%%%%%%%%%%%%%%%%%%%%%%%%%%%%%%%%%%%%
\subsection{Gravitational field}

In this section we discuss some of the geometrical properties of the gyratonic solutions.

One of the important characteristics of spacetimes are the scalar polynomial invariants
which are constructed only from the curvature and its covariant derivatives.
It was shown that gyratons in the Minkowski spacetime \cite{Fro-Is-Zel:2005:PHYSR4:}
have all the scalar polynomial invariants vanishing (VSI spacetimes) \cite{Prav-Prav:2002:CLAQG:},
the gyratons in the anti-de Sitter \cite{Fro-Zel:2005:PHYSR4:} and direct
product spacetimes \cite{Kadlecova:2009:PHYSR4:} have all invariants constant
(CSI spacetimes) \cite{Coley-Her-Pel:2006:CLAQG}.

In these cases, the invariants are independent of all metric functions
which characterize the gyraton, and have the same
values as the corresponding invariants of the background spacetime.
We observe that similar property is valid also for the gyraton on
Melvin spacetime, however, in this case the invariants are generally \emph{non-constant},
namely, they depend on the coordinate $\rho$. This property is a consequence of the general theorem
holding for the relevant subclass of the Kundt solution, see Theorem II.7 in \cite{ColeyEtal:2010}.

Values of some of the scalar curvature invariants can be found in Appendix \ref{apx:Invars}.

The metric \eqref{s5} admits the null vector ${k}=\partial_{v}$.
It is a Killing vector for ${g=0}$ and ${\div(\Sigma^2j)=0}$, i.e., for the no `heating' part in the gyratonic source.
The covariant derivative of ${k}$ is given by
\begin{equation}\label{recurrentk}
k_{\alpha;\beta}=-\Sigma^{-2}(\partial_{v}H) k_{\alpha}k_{\beta}+\Sigma^{-2}{k_{[\alpha}\nabla_{\beta]}\Sigma^2}\;.
\end{equation}
We observe that the congruence is not even recurrent \cite{Step:2003:Cam:}
as in the case of a gyraton on direct product spacetimes \cite{Kadlecova:2009:PHYSR4:}.
For $\partial_{v}H=0$ we recover the formula in Garfinkle and Melvin \cite{GarfinkleMelvin:1992:PHYSR4:}.
In general, the non-recurrency of the congruence is related to
the non-vanishing spin coefficient $\NP\tau=-\frac{1}{\sqrt{2}}\frac{1}{\Sigma^2}\Sigma_{,\rho}$,
cf.~\cite{Prav-Prav:2002:CLAQG:}, calculated in Appendix \ref{apx:NP}.

The null character of $k$ and the condition \eqref{recurrentk}
imply that the null congruence with tangent vector $k$ is
geodesic, expansion-free, sheer-free and twist-free, and the spacetime thus belongs
to the Kundt class.

Next we calculate components of the curvature tensors with respect to the
following adapted null tetrad ${\{k,\,l,\,m,\,\mb\}}$ \cite{Step:2003:Cam:}
\begin{equation}\label{b-vectors}
\begin{aligned}
  {k}&=\partial_{v}\;,\\
  {l}&=\frac{1}{\Sigma^2}(\partial_{u}-H\partial_{v})\;,\\
  {m}&=\frac{1}{\sqrt{2}\Sigma}\Bigl(a_{m}\,\partial_{v}
       +\partial_{\rho}-i\,\frac{\Sigma^2}{\rho}\partial_{\phi}\Bigr)\;.
\end{aligned}
\end{equation}
Here, we have introduced the projection of a transverse 1-form $a$ on the vector ${m}$
\begin{equation}
a_{m}= m^i a_i = a_\rho - i\,\frac{\Sigma^2}{\rho} a_\phi = (a +i\,{*a})_{\rho}\;,
%\quad a_\mb = \overline{a_m}\;,
\end{equation}
and we will use an analogous notation also for components of
the transverse gradient of a real function ${f}$
\begin{equation}
f_{,m}=m^i f_{,i} = f_{,\rho}-i\,\frac{\Sigma^2}{\rho}f_{,\phi}\;,\quad
f_{,\mb}=\overline{f_{,m}}\;.
\end{equation}
The dual tetrad of 1-forms ${\{\frm{k},\,\frm{l},\,\frm{m},\,\frm{\mb}\}}$ has a simple form
\begin{gather}\label{b-forms}
\frm{k}=\grad v + H\grad u-a\;,\quad
\frm{l}=\Sigma^2\grad u\;,\\
\frm{m}=\frac{\Sigma}{\sqrt{2}}\bigl(\grad \rho-i\,{*\grad \rho}\bigr)\;,\quad
\frm{\mb}=\frac{\Sigma}{\sqrt{2}}\bigl(\grad \rho+i\,{*\grad \rho}\bigr)\;.\notag
\end{gather}

With respect to this tetrad, we have found that the non-vanishing curvature components
are given by four new components and by those which are the same for the Melvin universe \eqref{o2}.
The non-vanishing  Ricci scalars are
\begin{equation}\label{sc3}
\begin{aligned}
\Phi_{12}
&=\frac{1}{4\sqrt{2}}\frac{1}{\Sigma^3}\Bigl(\Sigma^2 i b_{,m}+2ib(\Sigma^2)_{,\rho}+2g_{,m}\Bigr),\\
\Phi_{22}
&=\frac{1}{2}\frac{1}{\Sigma^2}\Bigl(\laplace H - (\Sigma^{-2})_{,\rho} H_{,\rho}+{\textstyle\frac12}\Sigma^2 b^2\\
&\qquad\qquad+2a^{i}g_{,i}+\Sigma^{-2}(g+\partial_{u})\div(\Sigma^2 a)  \Bigr)\;.\\
\end{aligned}
\end{equation}
and the non-vanishing Weyl scalars read
\begin{equation}\begin{split}\label{Weyl}
 \Psi_{3}&=\frac{1}{4\sqrt{2}}\frac{1}{\Sigma^3}\Bigl(\Sigma^2 i b_{,\mb}+ib(\Sigma^2)_{,\rho}+2g_{,\mb}\Bigr)\;,\\
 \Psi_{4}&=\frac{1}{2}\frac{1}{\Sigma^5}\biggl[2\Sigma\, a_{\mb}\,g_{,\mb}+\Sigma\biggl(H_{,\rho\rho}{-}\frac{\Sigma^4}{\rho^2}H_{,\phi\phi}{+}2i\frac{\Sigma^2}{\rho}H_{,\phi\rho}\biggr)\\
 &+\Sigma(g+\partial_{u})\left(a_{\rho,\rho}-\frac{\Sigma^4}{\rho^2}a_{\phi,\phi}+i\frac{\Sigma^2}{\rho}(a_{\rho,\phi}{+}a_{\phi,\rho})\right)\\
 &+\left(2\Sigma_{,\rho}-\frac{\Sigma}{\rho}\right)\biggl(H_{,\rho}+\partial_{u}a_{\rho}+g a_{\rho}\\
 &\qquad\qquad\qquad\qquad\quad +2i\frac{\Sigma^2}{\rho}(H_{,\phi}{+}\partial_{u}a_{\phi}{+}g a_{\phi})\biggr)\biggr]\;.\\
\end{split}\raisetag{48pt}
\end{equation}
In particular, there exists a relation between $\Psi_{3}$ and $\Phi_{12}$:
\begin{equation}\label{relpsiphi}
\overline{\Phi}_{12}+\Psi_{3}=\frac{1}{4\sqrt{2}}\frac{1}{\Sigma^3}\left[-i(\Sigma^2)_{,\rho}b+4g_{,\mb}\right].
\end{equation}

Therefore, the metric \eqref{s5} describes the transversal gravitational wave ($\Psi_{4}$ term) in the ${k}$ direction with
a longitudal wave component ($\Psi_{3}$ term). The gravitational wave is accompanied by an aligned pure
radiation field ($\Phi_{22}$ term) with non-null component ($\Phi_{12}$ term) propagating in the Melvin universe.
In fact, the scalars $\Psi_{3}$ and $\Phi_{12}$ are generated by the gyratonic functions $a_{i}$ and the function $g$.
In general the spacetime \eqref{s5} is of Petrov type $II$.

Let us now investigate the subcases of our solutions.
When we set the gyratonic functions $a_{i}=0$, the Ricci scalars become
\begin{equation}\begin{aligned} \label{ScRic}
\Phi_{12}&=\frac{1}{2\sqrt{2}}\frac{1}{\Sigma^3}g_{,m}\;,\\
 \Phi_{22}&=\frac{1}{2}\frac{1}{\Sigma^4\rho^2}\left[\rho(\rho H_{,\rho})_{,\rho}+\Sigma^4 H_{,\phi\phi}\right]\;\\
\end{aligned}\end{equation}
and the Weyl scalars then read
\begin{equation}\begin{aligned} \label{ScWey}
 \Psi_{3}&=\frac{1}{2\sqrt{2}}\frac{1}{\Sigma^3}g_{,\mb}\;,\\
 \Psi_{4}&=\frac{1}{2}\frac{1}{\Sigma^5\rho}\left[\Sigma\rho H_{,\rho\rho}+(2\rho\Sigma_{,\rho}-\Sigma)H_{,\rho}\right]\\
 &-\frac{1}{2}\frac{1}{\rho^2}H_{,\phi\phi}+i\frac{1}{2}\frac{1}{\Sigma^3\rho^2}\left[2\partial_{\rho}(\Sigma g H)-3\Sigma H\right]_{,\phi}.
\end{aligned}\end{equation}
We again obtain the spacetime with similar characteristics as for the full gyratonic  metric \eqref{s5}.
However, the scalars $\Psi_{3}$ and $\Phi_{12}$, which now depend only on function $g$, are now related by an even simpler relation:
\begin{equation}
\overline{\Phi}_{12}={\Psi}_{3}.
\end{equation}

If we assume additionally $H$ to be $v$-independent (i.e., $g=0$) we obtain the only
non-vanishing scalars $\Phi_{22}$ and $\Psi_{4}$ in the same form as in \eqref{ScRic} and \eqref{ScWey}.
This case and its subcases  were thoroughly discussed in \cite{Ortaggio:2004:PHYSR4:}.\footnote{%
The terms $\Phi_{22}$ and $\Psi_{4}$ have a little different form which is caused
by the slightly different choice of null tetrad in our paper.}

Finally, let us mention that for $\vrho=0$ the background reduces to Minkowski spacetime
and we thus recover the gyraton moving on the Minkowski background as an important subcase of our solutions.

%%%%%%%%%%%%%%%%%%%%%%%%%%%%%%%%%%%%%%%%%%%%%%%%%%%%%%%%%%%%%%%%%%%%%%%%%%%%%%%%%%%%
\subsection{Electromagnetic field}\label{ssc:elmag}

The gyraton propagates in a non-null electromagnetic field \eqref{EMF}, the influence of
which on the geometry is characterized by its density $\vrho$ \eqref{rhodef}.
The electromagnetic field is modified by the gyraton through the ${\sreal_i\grad u\wedge\grad x^i}$
terms. The electromagnetic field can be rewritten in terms of potentials using \eqref{sreal} as
\begin{equation}\begin{split}\label{realFS}
{F}&=E\bigl(\grad v \wedge\grad u+\grad u\wedge\grad \varphi\bigr)\\
&\quad+B\bigl(\Sigma^{-2}\rho\,\grad \rho\wedge\grad \phi+\grad u\wedge\grad \psi\bigr)\;.
\end{split}\end{equation}
It describes a superposition of electric and magnetic fields, both pointing
along the $z$ direction, which are modified by the gravitation field of the gyraton.
The additional term does not have a simple structure of electric or magnetic field,
however both are of the form ${\grad u\wedge\grad f}$ with ${f}$ being the proper potential.

The electromagnetic field  projected on the null tetrad \eqref{b-vectors}
is characterized by three scalars $\Phi_{i}$,
%\begin{equation}\begin{aligned}\label{EMPhiginv}
%&\Phi_{0}=0\;,\\
%&\Phi_{1}=\frac{\overline{\mathcal{B}}}{2\Sigma^2}\;,\\
%&\Phi_{2}=\frac{\overline{\mathcal{B}}}{\sqrt{2}\Sigma^3}\left[ a_{\mb}-\overline{\mathcal{S}}_{\rho}\right]\;.\\
%\end{aligned}\end{equation}
\begin{equation}\label{EMPhiginv}
\Phi_{0}=0\;,\quad
\Phi_{1}=\frac{\overline{\mathcal{B}}}{2\Sigma^2}\;,\quad
\Phi_{2}=\frac{\overline{\mathcal{B}}}{\sqrt{2}\Sigma^3}\left[ a_{\mb}-\overline{\mathcal{S}}_{\rho}\right]\;.\\
\end{equation}
It follows that the non-null electromagnetic field is aligned with the principal null direction ${k}$ of the gravitation field, but this vector is not a double degenerate vector of the field.

%%%%%%%%%%%%%%%%%%%%%%%%%%%%%%%%%%%%%%%%%%%%%%%%%%%%%%%%%%%%%%%%%%%%%%%%%%%%%%%%%%%%%%%%%%%%%
\section{Conclusion}\label{sc:conclusion}
%%%%%%%%%%%%%%%%%%%%%%%%%%%%%%%%%%%%%%%%%%%%%%%%%%%%%%%%%%%%%%%%%%%%%%%%%%%%%%%%%%%%%%%%%%%%%

We have derived and analyzed new gyraton solutions moving with the speed of light on
electro-vacuum Melvin background spacetime in four dimensions. This solution extends
the gyraton solutions previously known on the Nariai, anti-Nariai, and Pleba\'{n}ski--Hacyan
universes of type~D, and on conformally flat Bertotti--Robinson and Minkowski space.

The gyraton solutions describe a gravitational field created by a stress-energy tensor of
a spinning (circularly polarized) high-frequency beam of electromagnetic radiation,
neutrino, or any other massless fields. The gyratons generalize standard gravitational
{\it pp\,}-waves or Kundt waves by admitting a non-zero angular momentum of the source.
The interpretation is that the null matter in the interior of the gyratonic source
possesses an intrinsic spin (or non-zero angular momentum). This leads to other nontrivial
components of the Einstein equations, namely, ${G_{ui}}$ in addition to the pure
radiation $uu$-component which appears for {\it pp\,}-waves or Kundt waves.

We have shown that it is possible to define the gyraton by adding the gyratonic terms $a_{i}$
to the gravitational wave on the Melvin spacetime in cylindrical coordinates in a similar way
as we have defined them in the general Kundt class. We were able to find an ansatz for the
gyraton metric on the Melvin spacetime by direct transformation from the Kundt class of
metrics \eqref{ssc:def}.

We have further demonstrated that the Einstein--Maxwell equations reduce to the set of linear
equations on the 2-dimensional transverse spacetime which has a non-trivial geometry given
by the transverse metric. These equations can be solved exactly for any distribution
of the matter sources. In general, the problem has been thus reduced to a construction of
scalar Green functions for certain differential operators on the transverse space.

We have solved and analyzed the field equations for particular examples with the axial symmetry.
In these cases the equations reduce to ordinary differential equations.

We have analyzed geometric properties of the principal null congruence and we have found
that it is not recurrent contrary to the case of gyratons on direct product spacetimes.
We have explicitly calculated the curvature tensor and  determined that the gyratons on
Melvin spacetime are of Petrov type II and belong to the Kundt family of
shear-free and twist-free nonexpanding spacetimes.
The gyratonic term $a_{i}$ generates the non-trivial Ricci $\Phi_{12}$
and Weyl $\Psi_{3}$ scalars, in addition to the gravitational waves investigated in
\cite{Ortaggio:2004:PHYSR4:}. We found also a very simple
relation \eqref{relpsiphi} between these components. By studying particular subclasses
we have shown that our solutions are generalizations of those from \cite{Ortaggio:2004:PHYSR4:}.

The scalar polynomial invariants of the metric \eqref{s5} are in
general non-constant (although, some of them are zero)---they depend on the coordinate $\rho$.
The invariants are not affected by the presence of the gyratons,
they are the same as for the Melvin background. The same property was proved for
gyratons on backgrounds belonging to VSI or CSI families of spacetimes.

It would be interesting to investigate a generalization of our ansatz for more complicated
spacetimes which could allow, e.g., an inclusion of a cosmological constant.

%%%%%%%%%%%%%%%%%%%%%%%%%%%%%%%%%%%%%%%%%%%%%%%%%%%%%%%%%%%%%%%%%%%%%%%%%%%%%%%%%%%%%%%%%%%%
\begin{acknowledgments}
We wish to thank to Tom\'{a}\v{s} Pech\'{a}\v{c}ek and Otakar Sv\'{i}tek for helpful discussions
and to Marcello Ortaggio for his paper about gravitational waves in the Melvin universe which
motivated our work.
H.~K.  was supported by Grants No. GA\v{C}R-202/09/H033, No. GAUK~12209,
and  Project No. SVV~261301 of the Charles University in Prague.
P.~K. was supported by Grant No. GA\v{C}R~202/09/0772, and both authors thank  the
Project No. LC06014 of the Center of Theoretical Astrophysics.
\end{acknowledgments}

%%%%%%%%%%%%%%%%%%%%%%%%%%%%%%%%%%%%%%%%%%%%%%%%%%%%%%%%%%%%%%%%%%%%%%%%%%%%%%%%%%%%%%%%%%%%%
%%%%%%%%%%%%%%%%%%%%%%%%%%%%%%%%%%%%%%%%%%%%%%%%%%%%%%%%%%%%%%%%%%%%%%%%%%%%%%%%%%%%%%%%%%%%%
\appendix

%%%%%%%%%%%%%%%%%%%%%%%%%%%%%%%%%%%%%%%%%%%%%%%%%%%%%%%%%%%%%%%%%%%%%%%%%%%%%%%%%%%%%%
\section{The Einstein equations}\label{apx:AppA}
%%%%%%%%%%%%%%%%%%%%%%%%%%%%%%%%%%%%%%%%%%%%%%%%%%%%%%%%%%%%%%%%%%%%%%%%%%%%%%%%%%%%%%

Here we present quantities needed for evaluation of the Einstein equations.

The inverse to the metric \eqref{s5} is
\begin{equation}
\begin{split}\label{a0}
g^{\mu\nu}&\partial_{\mu}\partial_{\nu}=
  \frac{1}{\Sigma^2}\partial_{\rho}\partial_{\rho}+\frac{\Sigma^2}{\rho^2}\partial_{\phi}\partial_{\phi}
  -\frac{2}{\Sigma^2}\partial_{u}\partial_{v}\\
  &+2\bigl(a_{\rho}\frac{1}{\Sigma^2}\partial_{\rho}+a_{\phi}\frac{\Sigma^2}{\rho^2}\partial_{\phi}\bigr)\partial_{v}
  +(2\frac{H}{\Sigma^2}+a^2)\,\partial_{v}\partial_{v}.
\end{split}\raisetag{8ex}
\end{equation}

The stress-energy tensor $T^\EM$ of the electromagnetic field \eqref{EMF} can be defined as
\begin{equation}\label{defem}
T^\EM_{\mu\nu}=\frac{\epso}{2}{\mathcal{F}_{\mu}}^{\rho}\overline{\mathcal{F}}_{\nu\rho}
\end{equation}
where $\mathcal{F}\equiv {F}+i{{\star}F}$ is the complex self-dual Maxwell tensor.
The 4-dimensional Hodge dual is defined by
${\star}F_{\mu\nu}=\frac{1}{2}\varepsilon_{\mu\nu\rho\sigma}F^{\rho\sigma}$
and the Maxwell tensor ${{\mathcal{F}_{\mu\nu}}}$ satisfies the self-duality
condition  ${\star}\mathcal{F}=-i\mathcal{F}$.

The non-vanishing components of the stress-energy tensor \eqref{defem} are
\begin{align}
\varkappa T^\EM_{uv}&=\frac{\vrho}{\Sigma^2},\notag\\
\varkappa T^\EM_{uu}&=2\vrho\left(\frac{H}{\Sigma^2}+(\sslfdl-a)^2\right)\;,\notag\\
\varkappa T^\EM_{u\rho}&=\frac{\vrho}{\Sigma^2}(a_{\rho}-2\sslfdl_{\rho}),\notag\\
\varkappa T^\EM_{u\phi}&=\frac{\vrho}{\Sigma^2}(a_{\phi}-2\sslfdl_{\phi}),\notag\\
\varkappa T^\EM_{\rho\rho}&=\frac{\vrho}{\Sigma^2}=\frac{\vrho}{\Sigma^4}g_{\rho\rho}\;,\label{EMT}\\
\varkappa T^\EM_{\phi\phi}&=\frac{\vrho\rho^2}{\Sigma^6}=\frac{\vrho}{\Sigma^4}g_{\phi\phi}\;,\notag
\end{align}
where the density ${\vrho}$ was defined in~\eqref{rhodef}.

The non-vanishing components of the stress-energy tensor \eqref{m7} of the gyratonic matter are
\begin{equation}
\begin{gathered}
\varkappa T^\gyr_{uu}=j_u=v\,\div(\Sigma^2 j)+\iota\;,\\
\varkappa T^\gyr_{ui}=j_i\;.
\end{gathered}
\end{equation}

The Einstein tensor for the metric \eqref{s5} reads
\begin{align}
G_{uv}&= \frac{1}{\Sigma^2\rho}\bigl(-\rho(\Sigma_{,\rho})^2+2\Sigma(\Sigma_{,\rho})\bigr)\;,\notag\\
G_{uu}&=\frac{1}{2}\Sigma^4 b^2+\Sigma^2(\laplace H+\frac{(\Sigma^2)_{,\rho}}{\Sigma^4}H_{,\rho})
  +\Sigma^2(\partial^2_{v}H)a^2\notag\\
  &\quad +2\Sigma^2a^i\partial_{v}H_{,i}+(\partial_v H+\partial_{u})\,\div(\Sigma^2 a)\notag\\
  &\quad + 2HG_{uv}\;,\notag\\
G_{u\rho}&=\frac{1}{2}\frac{\Sigma^4}{\rho}b_{,\phi}-a_{\rho}\bigl(G_{uv}-\partial^2_v H\bigr)
    +\partial_{v}H_{,\rho}\;,\label{EinsteinT}\\
G_{u\phi}&=-\frac{1}{2}\rho(b_{,\rho}+\frac{4\Sigma_{,\rho}}{\Sigma}b)
    -a_{\phi}\bigl(G_{uv}-\partial^2_v H\bigr)+\partial_{v}H_{,\phi}\;,\notag\\
G_{\rho\rho}&=G_{uv}+\partial^2_v H\;,\notag\\
G_{\phi\phi}&=\frac{\rho^2}{\Sigma^4}(G_{uv}+\partial^2_v H)\;\notag.
\end{align}
%Another useful expressions are
%\begin{equation}\label{divrotident}
%\begin{aligned}
%\div (\Sigma^2 a)&=\Sigma^2\,\div a + \frac{\Sigma^2_{,\rho}}{\Sigma^2}\,a_{\rho}\;,\\
%\rot (\Sigma^2 a)&=\Sigma^2\,\rot a + \frac{\Sigma^2_{,\rho}}{\rho}\,a_{\phi}\;.\\
%\div (\Sigma^4 f)&=\Sigma^4\,\div f + 2(\Sigma^2)_{,\rho}f_{\phi\rho}\;,
%\end{aligned}
%\end{equation}
%which are used in simplifying the Einstein equations.
Here we have used only the metric \eqref{s5}, without any usage of the field equations.

%\bigskip\bigskip

%%%%%%%%%%%%%%%%%%%%%%%%%%%%%%%%%%%%%%%%%%%%%%%%%%%%%%%%%%%%%%%%%%%%%%%%%%%%%%%%%%%%%%
\section{The NP formalism}\label{apx:NP}

Calculating the Newman--Penrose spin coefficients with respect to the tetrad \eqref{b-vectors},
we recover again that the congruence ${k}$ is nonexpanding and
nontwisting (${\NP\rho=0}$), sheer-free (${\NP\sigma=0}$), geodesic and affine parameterized
(${\NP\kappa=\NP\eps=0}$). In addition, the tetrad is gauge invariant and it is not parallelly
transported along the null congruence because it does not satisfy ${\NP\kappa=\NP\pi=\NP\eps=0}$.

The remaining spin coefficients are
\begin{equation}\begin{gathered}\label{sc10}
\NP\lambda=0\;,\quad
\NP\mu=\frac{i}{2}b\;,\\
\NP\gamma=\frac{1}{4}\frac{1}{\Sigma^2}\bigl(2g+i\Sigma^2b)\;,\\
\NP\nu=\frac{1}{\sqrt{2}}\frac{1}{\Sigma^3}\left\{(g+\partial_{u})a_{\mb}+g_{,\mb}\right\}\;,\\
\NP\tau=-\frac{1}{\sqrt{2}}\frac{1}{\Sigma^2}\Sigma_{,\rho}\;,\quad
\NP\pi=+\frac{1}{\sqrt{2}}\frac{1}{\Sigma^2}\Sigma_{,\rho}\;,\\
\NP\alpha=\frac{1}{2\sqrt{2}}\frac{1}{\Sigma^2\rho}(2\rho\Sigma_{,\rho}-\Sigma)\;,\quad
\NP\beta=\frac{1}{2\sqrt{2}}\frac{1}{\Sigma\rho}\;.
\end{gathered}\end{equation}

%\pagebreak[2]

%%%%%%%%%%%%%%%%%%%%%%%%%%%%%%%%%%%%%%%%%%%%%%%%%%%%%%%%%%%%%%%%%%%%%%%%%%%%%%%%%%%%%%%
\section{Scalar polynomial curvature invariants}\label{apx:Invars}

%%%%%%%%%%%%%%%%%%%%%%%%%%%%%%%%%%%%%%%%%%%%%%%%%%%%%%%%%%%%%%%%%%%%%%%%%%%%%%%%%%

As we have already mentioned, the scalar curvature invariants are independent of all metric functions
which characterize the gyraton, and have the same values  as the corresponding invariants of the Melvin universe (cf.~\cite{ColeyEtal:2010}).
Let us stress here, however, that the invariants are generally \emph{non-constant},
namely, they depend on the coordinate $\rho$. In this appendix we list some of the curvature invariants.

The scalar curvature for the whole gyraton metric \eqref{s5} is zero, $R=0$.
Next, we define the following scalar polynomial invariants constructed
from the Riemann tensor:
\begin{equation}\begin{aligned}
R^{(2)}&=R^{ab}{}_{cd}R^{cd}{}_{ab}\;,\\
R^{(3)}&=R^{ab}{}_{cd}R^{cd}{}_{ef}R^{ef}{}_{ab}\;,\\
R^{(4)}&=R^{ab}{}_{cd}R^{cd}{}_{ef}R^{ef}{}_{pq}R^{pq}{}_{ab}\;,\\
R^{(5)}&=R^{ab}{}_{cd}R^{cd}{}_{ef}R^{ef}{}_{pq}R^{pq}{}_{rs}R^{rs}{}_{ab}\;.
\end{aligned}\end{equation}
Using the {\it GRtensor} package in {\it Maple}, we get the explicit expressions:
\begin{align}
R^{(2)}&=\frac{2\vrho^2}{\Sigma^8}\Bigl({\textstyle\frac38}\vrho^2\rho^4-3\vrho\rho^2+10\Bigr)\;,\notag\\
R^{(3)}&=-\frac{3\vrho^3}{\Sigma^{12}}\Bigl({\textstyle\frac1{16}}\vrho^3\rho^6
       -{\textstyle\frac34}\vrho^2\rho^4+7\vrho\rho^2-20\Bigr)\;,\notag\\
\begin{split}
R^{(4)}&=\frac{4\vrho^4}{\Sigma^{16}}\Bigl({\textstyle\frac9{256}}\vrho^4\rho^8
       -{\textstyle\frac{9}{16}}\vrho^3\rho^6\\
     &\qquad\qquad+{\textstyle\frac{51}8}\vrho^2\rho^4-33\vrho\rho^2+65\Bigr)\;,
\end{split}\label{invariants}\\
\begin{split}
R^{(5)}&=-\frac{5\vrho^5}{\Sigma^{20}}\Bigl({\textstyle\frac3{256}}\vrho^5\rho^{10}
       -{\textstyle\frac{15}{64}}\vrho^4\rho^8+{\textstyle\frac{31}{8}}\vrho^3\rho^6\\
     &\qquad\qquad-{\textstyle\frac{63}{2}}\vrho^2\rho^4+64\cdot127\vrho\rho^2-204\Bigr)\;.
\end{split}\notag
\end{align}
We explicitly observe that these invariants do not depend on any of the metric
functions $a_{i}$, ${g}$, and ${h}$ which characterize the gyraton.

The invariants \eqref{invariants} mimic the behavior of the Gauss curvature of the
transverse space discussed in detail in \eqref{ssc:transsp}.
They have their maximum on the axis $\rho=0$ and they are vanishing as
``the neck of the vase closes off asymptotically'' as $\rho$ tends to infinity.
For $\vrho=0$ we get the identically vanishing invariants,
i.e., the invariants for the gyratons on Minkowski background (VSI).

In {\it Maple} tensor package {\it GRtensor} there is defined a set of curvature invariants \texttt{CMinvars}.
%\texttt{CMinvars} = $\{R,\, R_{1},\, R_{2},\, R_{3},\, W1R,\, W1I,\, W2R,\, W2I,\, M1R,\, M2I,\,$ $M3,\, M4,\, M5R,\, M5I\}$.
For completeness, we present the explicit expressions for them:
\begin{equation}\begin{gathered}
R=R_{2}={W1I}={W2I}=0\;,\\
R_{1}=\frac{\vrho^2}{\Sigma^8}\;,\quad{W1R}=\frac{3\vrho^2}{2^5\Sigma^{8}}(\vrho\rho^2-4)^2\;,\\
R_{3}=\frac{\vrho^4}{4\Sigma^{16}}\;,\quad{W2R}=\frac{3\vrho^3}{2^8\Sigma^{12}}(\vrho\rho^2-4)^3\;,\\[1ex]
{M1I}={M2I}={M4}={M5I}=0\;,\\
{M1R}=\frac{\vrho^3}{2^2\Sigma^{12}}(\vrho\rho^2-4)\;,\\
{M2R}={M3}=\frac{\vrho^4}{2^4\Sigma^{16}}(\vrho\rho^2-4)^2\;,\\
{M5R}=\frac{\vrho^5}{2^{6}\Sigma^{20}}(\vrho\rho^2-4)^3\;.
\end{gathered}\end{equation}\\[-12pt]

%\vfill

%%%%%%%%%%%%%%%%%%%%%%%%%%%%%%%%%%%%%%%%%%%%%%%%%%%%%%%%%%%%%%%%%%%%%%%%%%%%%%%%%%
%%%%%%%%%%%%%%%%%%%%%%%%%%%%%%%%%%%%%%%%%%%%%%%%%%%%%%%%%%%%%%%%%%%%%%%%%%%%%%%%%%

%\bibliography{Hedvika}% Produces the bibliography via BibT

\end{document}